\documentclass[reprint,amsmath,amssymb,floatfix,aps]{revtex4-2}
\usepackage[version=3]{mhchem} % Formula subscripts using \ce{}
\usepackage{graphicx}
\usepackage{enumitem}
\usepackage{xcolor}
\usepackage[normalem]{ulem}
\usepackage[fontsize=10pt]{scrextend}
\usepackage{graphicx}
\usepackage{wrapfig}
\usepackage{framed}
\usepackage{color}
\usepackage{float}
\usepackage{multirow}
\usepackage{amsmath} 
\makeatletter

\newcommand*{\rom}[1]{\expandafter\@slowromancap\romannumeral #1@}
\makeatother

\begin{document}
\preprint{APS/123-QED}
%%%%%%%%%%%%%%%%%%%%%%%%%%%%%%%%%%%%%%%%%%%%%%%%%%
\title{Motility and pair-wise interactions of chemically active droplets in 1-D confinement}
%%%%%%%%%%%%%%%%%%%%%%%%%%%%%%%%%%%%%%%%%%%%%%%%%%
\author{Pawan Kumar}
\author{Prateek Dwivedi}
\author{Sobiya Ashraf} 
\author{Dipin Pillai}
\author{Rahul Mangal}
\email{mangalr@iitk.ac.in}
\affiliation{Department of Chemical Engineering, Indian Institute of Technology Kanpur, Kanpur, India.}

%%%%%%%%%%%%%%%%%%%%%%%%%%%%%%%%%%%%%%%%%%%%%%%%%%%%%%%%%%%
\begin{abstract}
 Self-propelled droplets serve as ideal model systems to delve deeper into understanding of the motion of biological micro-swimmers by simulating their motility. Biological microorganisms are renowned for showcasing a diverse array of dynamic swimming behaviors when confronted with physical constraints. This study aims to elucidate the impact of physical constraints on swimming characteristics of biological microorganisms. To achieve this, we present observations on the individual and pair-wise behavior of micellar solubilized self-propelled 4-Cyano-4'-pentyl-biphenyl (5CB) oil droplets in a square capillary channel filled with a surfactant trimethyl ammonium bromide (TTAB) aqueous solution. To explore the effect of the underlying Péclet ($Pe$) number of the swimming droplets, the study is also performed in the presence of additives such as high molecular weight polymer Polyethylene oxide (PEO) and molecular solute glycerol. The capillary confinement restricts droplet to predominantly one-dimensional (1D) motion, albeit with noticeable differences in their motion across the three scenarios. Through a characterization of the chemical and hydrodynamic flow fields surrounding the droplets, we illustrate that the modification of the droplets' chemical field due to confinement varies significantly based on the underlying differences in the Péclet number ($Pe$) in these cases. This alteration in the chemical field distribution notably affects the individual droplets' motion. Moreover, these distinct chemical field interactions between the droplets also lead to variations in their pair-wise motion, ranging from behaviors like chasing to scattering.

\end{abstract}
%%%%%%%%%%%%%%%%%%%%%%%%%%%%%%%%%%%%%%%%%%%%%%%%%%%%%%%

\maketitle

\section{Introduction}
To execute efficient self-propulsion in different viscous environments, microorganisms utilize non-reciprocity by employing local asymmetry to self-propel at low Reynolds number \cite{purcell1977life}. They utilize appendages attached to their surface or deform their bodies to achieve non-reciprocal strokes that introduce the required symmetry breaking in their mechanical strokes. These motile microorganisms also possess the remarkable ability to sense and respond to their surroundings and appropriately respond to the external hydrodynamic and chemical fields in their vicinity. For example, it has been observed that \textit{E. coli} cells demonstrate faster movement in viscoelastic media \cite{zottl2019enhanced}. Similarly, Euglena cells are known to adapt to flagellar swimming in a confined channel without contacting the wall, and to crawling as confinement increases, including rounding and peristaltic cell deformations \cite{noselli2019swimming}. In an effort to mimic and gain deeper insights into the propulsion strategies employed by microorganisms, researchers have designed and fabricated various artificial micro-motors at the laboratory scale, that include chemically powered Janus micro-/nano-motors and interfacial tension-driven water/oil droplets in oil/water. The advancement and understanding of micro-motors also hold tremendous promise for a wide range of futuristic applications, such as targeted drug delivery \cite{xu2018sperm}, absorbents for chemical spills \cite{gao2014environmental}, and bio-sensors \cite{jurado2017janus}.

The autonomous motion of Janus particles (JPs) is achieved through their preferential asymmetric reactivity with the surrounding media, which relies on self-diffusiophoresis (chemical gradient) \cite{wurger2015self, zhou2018photochemically}, self-thermophoresis (temperature gradient) \cite{jiang2010active,cao2021photoactivated}, or self-electrophoresis (electric potential gradient)\cite{dou2016directed, gangwal2008induced} mechanisms. Active droplets (ADs), on the other hand, do not possess inherent chemical asymmetry and therefore rely on spontaneous asymmetry generating mechanisms to generate self-propulsion, such as micellar solubilization \cite{dwivedi2021solute,izzet2020tunable,izri2014self,meredith2022chemical,peddireddy2012solubilization}, phase separation \cite{thakur2006self}, and reaction-diffusion processes \cite{kumar2021fast}. 

Among all, the micellar-solublization based propulsion mechanism is considered to be the simplest. The solubilization of the droplet in the form of nano-emulsions/filled micelles, generates a spontaneous heterogeneity in the distribution of filled and empty micelles surrounding the droplet, resulting in an interfacial tension gradient. The resulting Marangoni stresses drive the surrounding fluid flow toward regions of higher surface tension, causing the droplet to move in the opposite direction. The competition between the advective and diffusive time scales of the solutes around the droplet governed by the nonlinear coupling of hydrodynamics and advection-diffusion processes plays a crucial role in governing the swimming mode \cite{morozov2019nonlinear}. With increase in Péclet number ($Pe$) Dwivedi \textit{et al.} recently demonstrated a transition from persistent to random to a jittery motion \cite{dwivedi2023mode}. Suda \textit{et al.} also reported a straight to curvilinear motion with increase in $Pe$ \cite{suda2021straight}. Over the years, several studies have reported intriguing aspects of the isolated motion of these droplets, including nematic elasticity induced curling \cite{kruger2016curling, suga2018self}, solute induced jitteriness \cite{dwivedi2021solute, hokmabad2021emergence},  auto-negative chemotaxis \cite{jin2017chemotaxis}, upstream rheotaxis in external flow \cite{dwivedi2021rheotaxis, dey2022oscillatory}, deformation in viscoelastic medium \cite{dwivedi2023deforming},  and more. Some intriguing multi-body behavioral aspects have also been reported such as predator-prey behavior \cite{meredith2020predator}, the emergence of spontaneously rotating clusters \cite{hokmabad2022spontaneously}, dynamic-assemblies \cite{thutupalli2018flow} and chemotactic self-caging \cite{hokmabad2022chemotactic}. 

Most studies of ADs have been carried out in a Hele--Shaw cell, offering a 2D confined space. However, confinement to 1D domains and the interaction with the surrounding walls can influence swimmers' dynamics and collective behavior by virtue of hydrodynamics and chemical field alteration. Biological microorganisms often navigate through narrow channels and porous media, and their mobility patterns and navigation strategies have been shown to be affected by the surrounding environments. Theoretical studies have examined the propulsion of flagella in cylindrical tubes, considering helical swimmers \cite{liu2014propulsion,acemoglu2014effects}, and the swimming mode transition (i.e. pusher or puller) of amoeboid in parallel wall confinement \cite{wu2015amoeboid}. 
For phoretic micro-swimmers in close proximity to a solid wall, theoretical and experimental investigations have reported behavior including rebounding, scattering, hovering, and sliding \cite{mozaffari2016self,uspal2015self,ibrahim2016walls, desai2022steady}. For ADs, theoretical work by Desai \textit{et al.} \cite{desai2021instability} reported the accumulations of chemical solutes and subordinate effects of hydrodynamics, at low Péclet number, inducing swimming away from the wall. However, with an increase in $Pe$, advection of fluid reduces the distribution of chemical solute at the frontal region of droplet, causing the droplet to approach close to the wall \cite{lippera2020collisions}. Recent experimental work by de Blois \textit{et al.} has demonstrated that a swimming droplet (water-in-oil) reduces its speed by increasing confinement, and even splits into fragments under the strong confinement of the capillary \cite{de2021swimming}. Despite these recent explorations, understanding of the complex behaviors of swimmers in quasi 1D confinements is still in its infancy. In particular, for ADs, where the nonlinear coupling of chemical and hydrodynamics fields introduce additional complexity remains mostly unexplored. 

In this article, we have investigated the individual and pair-wise dynamics of micellar solubilization based self-propelled oil droplets in a square capillary offering quasi 1-D confinement. The associated advection-diffusion effects are varied by adding polymer/glycerol as additives to the continuous aqueous phase. In-situ fluorescence visualization of the trail of filled micelles and the fluid-flow field around the droplets provide a better understanding of the underlying droplet motion. Understanding swimming behavior of ADs at the laboratory scale in confined environments can provide crucial insights into biological processes and pave the way for technological applications in microfluidics \cite{son2015live}.  

\section{Materials and Methods}

Using a micro-injector (Femtojet 4i, Eppendorf), 4-Cyano-4'-pentyl-biphenyl (5CB, Jiangsu hecheng advanced materials Co., Ltd.) oil droplets of diameter 330 $\pm$ 25 $\mu$m were injected in an aqueous solution of 6 wt.$\%$ trimethyl ammonium bromide (TTAB, Loba Chemicals), used as surfactant. For different experiments, 1.125 wt.$\%$ of high molecular weight (8000 kDa) polyethylene oxide (PEO, Sigma Aldrich) and 80 wt.$\%$ glycerol (Loba chemicals) were added separately to the aqueous solution. To perform the experiments in a quasi 2-dimensional setting, a custom-made Hele--Shaw cell (see schematic shown in figure 1(A)), with dimensions (1.5 $\times$ 1.5 cm$^2$) and thickness $\sim$ 500 $\mu$m was fabricated using clean glass slides. For the quasi 1-dimensional experiments, a square glass capillary of edge length 500 $\mu m$ and axial length of 4 to 10 cm was used ({see schematic shown in figure 1(B)}). The droplet laden solution was injected into the optical cell/capillary. To maintain constant temperature at 25 $^o$C, the imaging chambers were mounted on a thermal stage on the upright microscope, Olympus BX53. The self-propelled motion of droplets was recorded with Olympus LC30 or BFS-U3-70S7CC, FLIR, camera at 1-30 frames s$^{-1}$ in the bright-field mode.
\begin{figure}[t]
  {\includegraphics[scale=0.3]{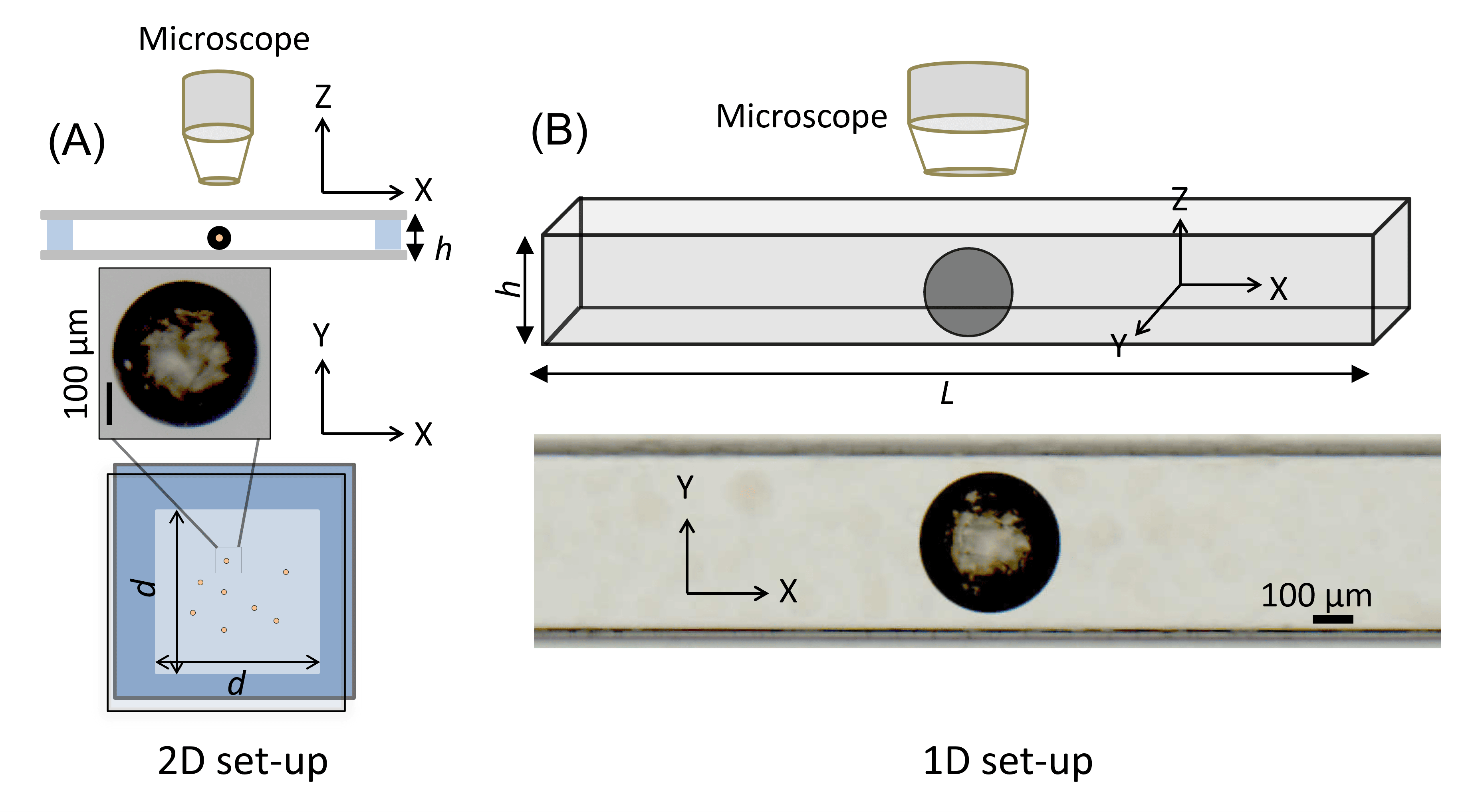}}
\caption{\small (A) Schematic of the quasi 2-dimensional experimental setup where a Hele-shaw cell is used as the optical chamber with $h$ = 500 $\mu$m. The corresponding optical micrograph shown in the inset corresponds to a 5CB swimming droplet in just an aqueous TTAB solution. (B) Schematic of the square capillary channel used for quasi 1-dimensional experiments and corresponding optical micrograph illustrating a 5CB swimming droplet in just aqueous TTAB solution.}
\label{fig1}
\end{figure}

Using fluorescent polystyrene particles (3.2 $\mu$m) as tracers, fluid flow around the self-propelled droplets was characterized using particle image velocimetry (PIV) experiments. To probe the chemical field of filled micelles around the droplets, an oil-soluble fluorescent dye (Nile Red, Sigma Aldrich) was added to the 5CB phase. Further, the cell was mounted on an inverted microscope Olympus IX73 equipped with a fluorescence illuminator Olympus U-RFL-T (Mercury Burner USH-1030L), with laser ($\lambda$ $\sim$ 560 nm) to excite the dye molecules. ORX-10G-71S7C-C, FLIR, camera connected with the microscope (4-8 X magnification), was used to record the fluorescence videos. The $x-y$ trajectories were characterized by tracking the centroid of droplet with the Image-J software using Mtrack2 plugins. The velocity vector flow field around the droplets was characterized using PIVlab open-source software in MATLAB \cite{thielicke2021particle}, and streamlines were plotted in Techplot 360.

\begin{figure*}[t]
  {\includegraphics[scale=0.57]{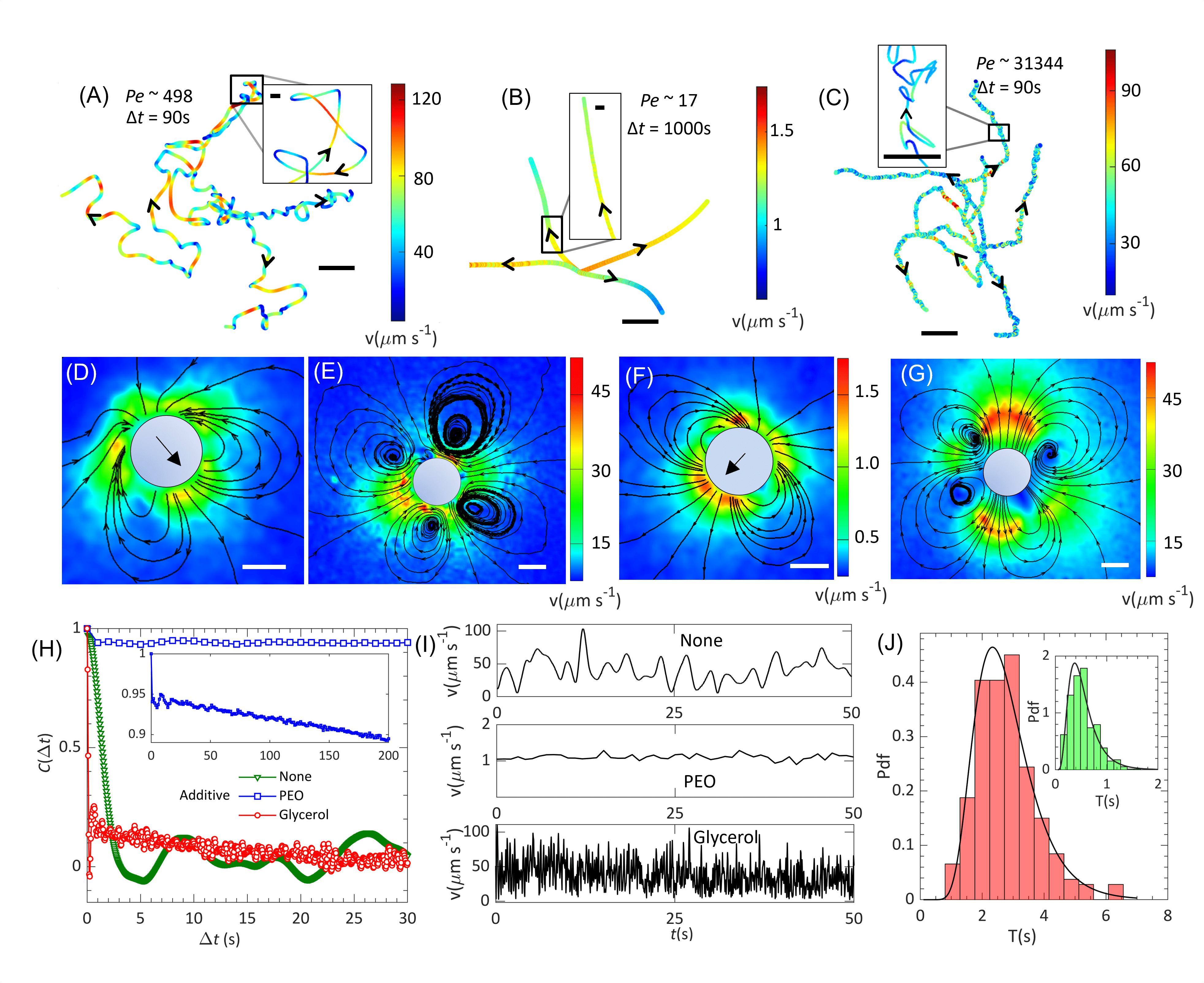}}
\linespread{1.0}
\caption{\small Representative $x-y$ trajectories captured for duration $\Delta${\textit{t}}, of the centroids of 5CB swimming droplets in TTAB aqueous solutions containing (A) no additional solute, (B) PEO and (C) Glycerol. Scale bars in (A)-(C) correspond to 350 $\mu$m. Scale bars in the insets correspond to 20 $\mu$m. Corresponding PIV micrographs with flow streamlines around the swimming 5CB droplet in (D, E), (F) and (G), respectively. Scale bar = 200 $\mu$m (H) Corresponding temporal evolution of the angular autocorrelation functions. (I) Speed dynamics of a 5CB droplet in TTAB aqueous solutions containing none, PEO and glycerol additives. (J) Histogram plots of inter-peak time difference for TTAB aqueous solution (red) and glycerol (inset). }
\label{fig2}
\end{figure*}

\section{Results and discussion}

\subsection{Dynamics in Quasi 2D system}

First, we investigate the behavior of active 5CB droplets ($\sim$ 330 $\mu$m) in 6 wt.$\%$ TTAB aqueous solution, with and without glycerol or PEO as additives, in a quasi-2D confinement. As a consequence of the bulk concentration of TTAB surpassing its critical micelle concentration (CMC) in all of the solutions, the droplets undergo micellar solubilization, leading to their self-propelled motion. \cite{dwivedi2022self, peddireddy2012solubilization} As the droplets move, they leave a trail of filled micelles behind them. The interplay between the advection and diffusion of micelles around the droplets plays a crucial role in determining their swimming behavior. The occurrence of spontaneous self-propulsion, driven by the interfacial instability, is contingent upon the Péclet number $Pe = \frac{R \langle \text{v} \rangle}{D}$, where $\langle$v$\rangle$ denotes the droplet average propulsion speed in surfactant media, $R$ represents the droplet size, and $D$ is the diffusion coefficient of the micelles or surfactant monomer \cite{morozov2019nonlinear}. Here, $Pe$ is determined through experimental characterization of average droplet speed $\langle \text{v} \rangle = \langle{|{\frac{\textbf{r}_{i+1} - \textbf{r}_i}{t_{i+1} - t_i}}|} \rangle$, where \textbf{r$_{i}$}($x_{i}$,$y_{i}$) is the instantaneous position vector, and micelle diffusivity $D$ values have been adopted from the previous experimental work \cite{dwivedi2023mode}.

Figure \ref{fig2}(A) illustrates the representative $x-y$ trajectories demonstrating the motion of 5CB swimming droplets in 6 wt.$\%$ TTAB aqueous solution with no additional solute. The unbiased nature of trajectories indicates the absence of any external convection within the system. While $\langle \text{v} \rangle$ was measured to be $\sim$ 42 $\mu$m s$^{-1}$, the color coding depicts the instantaneous speed v$_{t}$. While the droplets slow down during turns, they propel faster during the persistent part of their trajectory, reminiscent of a run and tumble like motion. The inset illustrates a zoomed-in view of a section of the trajectory, highlighting the direction fluctuations. The characterization of the flow field illustrated in figure \ref{fig2}(D) around the droplet's vicinity demonstrates a dipole formation during the droplet's motion, and fitting the tangential velocity using the squirmer model reveals a pusher mode of swimming ($\beta$ = -0.7). Moreover, a quadrupole mode (figure \ref{fig2}(E)) is observed during the droplet's halting state. These observations of jitteriness in the motion are consistent with the associated higher $Pe \sim 498$. When a high molecular weight polymer (PEO 8000 kDa) is introduced to the TTAB aqueous solution, it not only increases the bulk viscosity but also induces viscoelastic properties in the medium. In such macromolecular aqueous solutions where the surfactant/micelle size is smaller than the host polymer size, the surfactant/micelle diffusivity has been recently identified to be unchanged and equal to the diffusivity in the solvent, i.e., water \cite{dwivedi2023mode}. Due to the combination effect of slower speed ($\sim$ 1.24 $\mu$m s$^{-1}$) and higher than expected micelle diffusivity (24 $\mu$m$^{2}$ s$^{-1}$), the swimming droplets manifest lower $Pe$ ($\sim$ 17) values resulting in smooth persistent motion. A few representative $x-y$ trajectories of swimming 5CB droplets in the presence of high polymer concentration ($c_\text{PEO}$) are presented in figure \ref{fig2}(B). Fitting the peripheral flow field around the droplet with the squirmer model results in $\beta$ $\sim$ -0.4, a signature of the droplet being a weak pusher (see supporting figure S1). When 80 wt.$\%$ glycerol is added to the continuous phase, due to an increase in bulk viscosity ($\sim$ 40 times), the diffusion coefficient of micelles decreases, resulting in nearly 57-fold increase in the Péclet number ($\sim$ 31344) compared to the former case of pure aqueous solution of TTAB. The representative $x-y$ trajectories shown in figure \ref{fig2}(C) exhibit enhanced jitteriness, while the flow field (figure \ref{fig2}(G)) also demonstrates the quadrupole modes at the halts. These observations are consistent with previous studies conducted in the presence of glycerol in TTAB solution \cite{dwivedi2021solute,hokmabad2021emergence}. 

It is to be noted that in contrast to our previous work \cite{dwivedi2023mode}, the droplet size in our current study is about 4 times larger. With an increase in $Pe$, a transition from pusher to quadrupole-dipole (bistable) to dominant quadrupole mode is observed. Nevertheless, the transitions are consistent with the $Pe$-based phase-diagram proposed earlier.

Furthermore, the corresponding angular autocorrelation function 
    $\textit{C}(\Delta t) = \langle \frac{\textbf{\text{v}}(\Delta t).\textbf{\text{v}}(0)}{|\textbf{\text{v}}(\Delta t)||\textbf{\text{v}}(0)|} \rangle $
 of the trajectories in different solutions is shown in figure 2(H). Here, $\langle ... \rangle$ demonstrates the time average over $\Delta t$. In the case of droplet motion in a pure aqueous TTAB solution, the angular autocorrelation diminishes to zero and displays oscillations with diminishing amplitude over time. This decline in $C(\Delta t)$ to zero affirms that the droplet's motion is non-directional. The oscillations around zero indicate a slight curling behavior in the droplet trajectories, which is attributed to the nematic phase of the droplet \cite{kruger2016curling}. In the PEO solution, which corresponds to lower $Pe$ values, $C(\Delta t)$ does not exhibit any noticeable decay, confirming the long-lasting and persistent nature of the trajectories. Over longer time intervals, $C(\Delta t)$ gradually diminishes to zero, highlighting the slow evolution of randomness in the long-term trajectories. However, in the presence of glycerol, there is an increased jitteriness in the droplet's motion. For small $\Delta t$ values, $C(\Delta t)$ rapidly decreases to zero, emphasizing the short-term randomness in the trajectories.  Figure \ref{fig2}(I) depicts the temporal variation of droplet speed in the three cases. In the presence of PEO, the droplet maintains nearly constant speed, while in the other two cases, fluctuations in speed with time are evident. Figure \ref{fig2}(J) illustrates the corresponding probability distribution of the inter-peak time difference (T) of the fluctuations. Clearly, oscillations occur more rapidly in the presence of glycerol (as indicated in the inset) compared to the case of the pure TTAB aqueous solution.
 
\subsection{Dynamics in a capillary}

\begin{figure*}[t]
  {\includegraphics[scale=0.58]{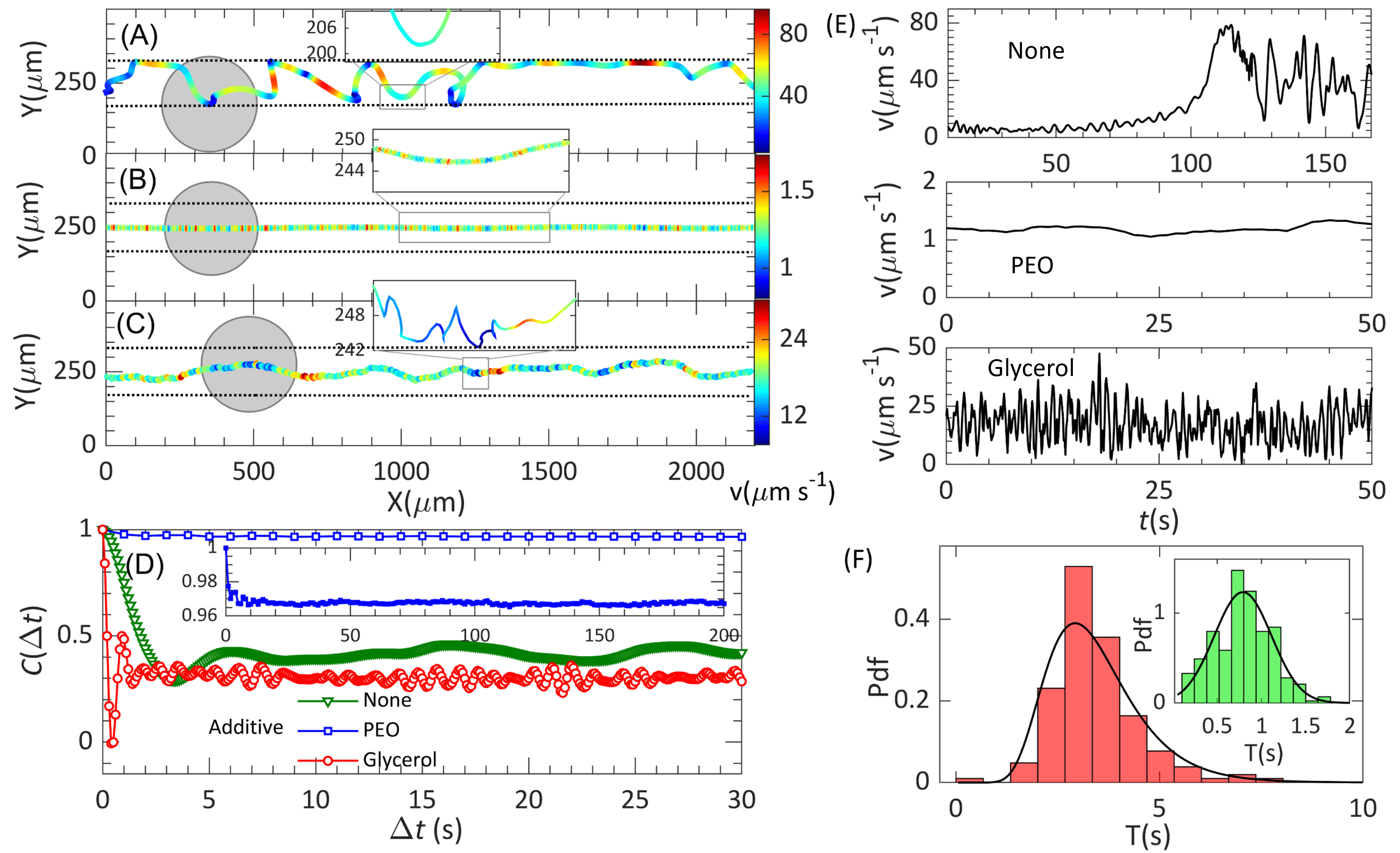}}
\linespread{1.0}
\caption{\small Representative $x-y$ trajectories of the centroids of 5CB swimming droplets in aqueous TTAB solutions containing (A) nothing (B) PEO and (C) Glycerol. The dotted lines represent the lateral boundaries for the droplet centroid. Insets in the figures illustrate the zoomed-in section of the trajectory. (D) Corresponding evolution of the angular autocorrelation function. (E) Speed evolution in different TTAB aqueous solutions with and without additives. (F) Histogram plots of inter-peak time difference of speed time-series for TTAB aqueous solution (red) and glycerol (inset). }
\label{fig3}
\end{figure*}

After verifying the expected behavior of swimming droplets in quasi-2D environment, we set out to conduct experiments in a square glass capillary offering a quasi-1D swimming, see schematic illustration shown in figure \ref{fig1}(B). Upon introducing a 5CB droplet ($\sim$ 330 $\mu$m) from one end of the capillary, within a few seconds, droplet commences its active motion. Since the capillary cross-section is marginally larger compared to the droplet size, the droplet remains spherical (see optical micrograph shown in figure \ref{fig1}(B)). However, the lateral confinement limits the droplet motion mostly along the axial (i.e. +$x$) direction, with varying dynamical attributes in different surroundings, which we discuss in detail below.

For a swimming droplet in just the aqueous TTAB solution, a representative trajectory shown in figure \ref{fig3}(A) illustrates a relatively complex path traced by the droplet (see representative supporting movie S1). In part of the trajectory (0 $\mu$m $\leq$ $x$ $\leq$ 1200), while the droplet moves along the positive $x$-axis, it exhibits irregular undulations resulting from sharp and smooth turns along the $y$-axis. 
In another section of the capillary (1300 $\mu$m $\leq$ $x$ $\leq$ 1900), the droplet appears to move along the wall resulting in a nearly linear trajectory. Nevertheless, in contrast to the two-dimensional (2D) system, where motion was primarily random, the quasi one-dimensional (1D) confinement enforced by the capillary predominantly directs the droplet's net motion along the capillary axis. In the presence of PEO, the representative $x-y$ trajectory shown in figure \ref{fig3}(B) demonstrates that droplet exhibits an unexpected straight line motion along the capillary axis (+$x$-axis) (see representative supporting movie S2). Closer inspection of the trajectory reveals the existence of minor speed fluctuations along with minor changes in direction. In the presence of 80 wt.$\%$ glycerol, as indicated by the representative $x-y$ trajectory shown in figure \ref{fig3}(C), droplet's long time motion is observed to be undulating about the capillary axis with varying lateral gap with the walls (see representative supporting movie S3). A closer inspection of the trajectory reveals the existence of speed and spatial fluctuations associated with the short time jittery motion. A comparison of the droplets' swimming speed and the resulting $Pe$ values have been tabulated in table \ref{table}. 

\begin{table*} 
   \caption{Characteristics of droplet motion in aqueous TTAB solution doped with different additives. Here, subscript $w$ denotes the droplet motion along the wall while $c$ represents motion away from the wall.}
   \setlength{\tabcolsep}{0.7\tabcolsep}% Shrink \tabcolsep by 30%
    \centering
    \begin{tabular}{c c c c c c c c}
    \hline
    Additive & \multicolumn{3}{c}{$\langle \text{v} \rangle$} &  $D$ & \multicolumn{3}{c} {$Pe$}\\
     & \multicolumn{3}{c}{$\mu$m s$^{-1}$} & $\mu$m$^{2}$ s$^{-1}$ & \multicolumn{3}{c}{}\\
    \hline
    & 1D$_{w}$ & 1D$_{c}$ & 2D & &1D$_{w}$ & 1D$_{c}$ & 2D\\ 
    \hline 
    None  & 12$\pm$6 & 37$\pm$3 & 42$\pm$4 & 29 &140$\pm$70& 431$\pm$23 & 498$\pm$26 \\
    PEO  &-- &1.24$\pm$0.27 &1.20$\pm$0.1 & 24 &--& 18$\pm$4 &17$\pm$1\\
    Glycerol  & --&18$\pm$2 &37$\pm$3 &0.4 &--& 14617$\pm$1855 &31344$\pm$3416\\
       \hline
    \end{tabular}
    \label{table}
     \end{table*}

The characteristics of the droplet motion in the capillary while swimming in varying surroundings are also reflected in their angular auto-correlation data shown in figure \ref{fig3}(D). In just the aqueous TTAB solution, due to quasi 1D confinement which restricts the motion of the droplet along +$x$ axis, $C(\Delta t)$ does not fully decay to zero. Although, the irregular undulations in the trajectory cause some oscillations in $C(\Delta t)$, the lack of any negative values confirms that the droplet motion is mostly in the forward direction with negligible backward steps. In case of PEO aqueous solution, the short time minor direction fluctuations of the droplet lead to a discernible decay in $C(\Delta t)$ at short $\Delta t$, as depicted in zoomed inset figure \ref{fig3}D. However, in contrast to the 2D system, $C(\Delta t)$ does not demonstrate any decay at long time scales, indicating a persistence motion along the capillary axis. For droplet motion in glycerol aqueous solution, droplet's jittery dynamics closely resemble that of a 2D system, causing $C(\Delta t)$ to sharply decay to zero at short $\Delta t$. Nevertheless, as illustrated in figure \ref{fig3}(D), due to capillary confinement restricting droplet motion mostly along the capillary axis, at longer time scales $C(\Delta t)$ saturates $\sim$ 0.3. Figure \ref{fig3}(E) demonstrates the corresponding temporal evolution of droplet speed in the three cases. Within the aqueous TTAB solution, it is noticed that the amplitude of speed fluctuations of the droplet is minimal as it moved along the wall. However, these fluctuations significantly increases once the droplet detaches from the wall. The histogram of the corresponding inter-peak time difference (T) of the speed time-series, shown in figure \ref{fig3}(F), demonstrates a uni-modal fit with a mode at 2.94 s, indicating no significant variation in the time-period of speed oscillations. In fact, the peak of the distribution remains nearly close to the 2D system ({2.35 s}). In presence of glycerol, it appears that due to the 1D confinement, compared to the 2D case ($\langle v \rangle$ $\sim$ 37 $\mu$m s$^{-1}$), the droplet speed reduces ($\langle v \rangle$ $\sim$ 18 $\mu$m s$^{-1}$). However, as shown in the inset of figure \ref{fig3}(F), the mode of the distribution of inter-peak of speed time-series increases $\sim$ 0.8 s marginally when compared to the motion in the 2D system ($\sim$ 0.37 s). In the presence of PEO solution, similar to the 2D system, the droplet exhibits motion with a nearly constant speed. 

\begin{figure*}[ht!]
  {\includegraphics[scale=0.55]{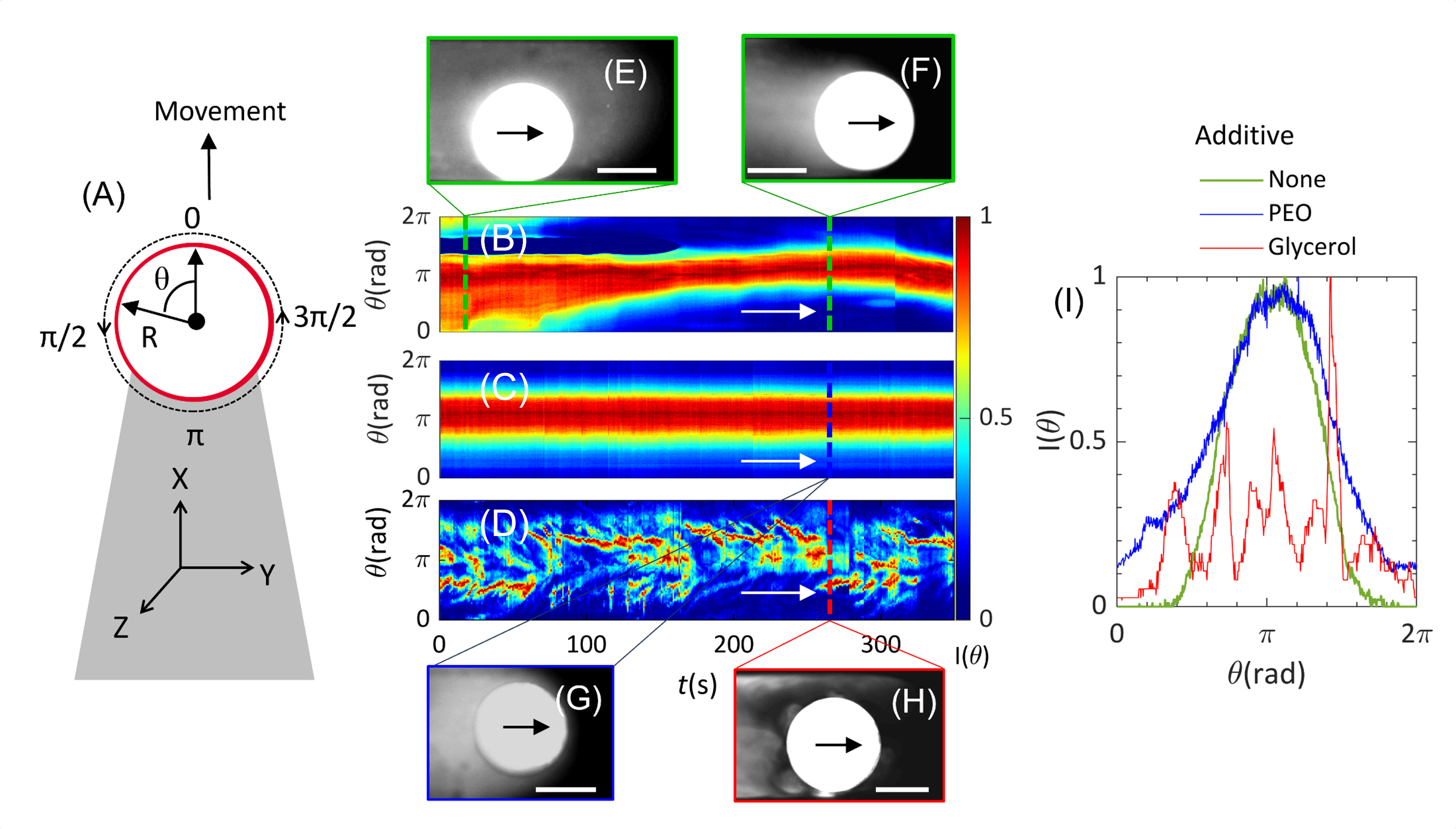}}
\linespread{1.0}
\caption{\small (A) Schematic to characterize the distribution of chemical field around the droplet. Kymograph plots are obtained for swimming 5CB droplets in aqueous TTAB solutions containing (B) nothing (C) PEO and (D) Glycerol. Corresponding grayscale fluorescence micrographs of swimming 5CB droplet releasing fluorescent plumes of chemical field are shown in (E, F), (G) and (H), respectively. Scale bar = 200 $\mu{m}$. (I) Variation of normalized fluorescence intensity plots of chemical plumes around the droplet.}
\label{fig4}
\end{figure*}

To understand the dynamics of self-propelled droplets within the 1D constraint of a capillary and discern the distinctions from their movement in a 2D setting, we conducted a thorough analysis of the spatio-temporal evolution of the chemical field (filled micelles) released by the droplets in their vicinity. As illustrated in the schematic \ref{fig4}(A), the peripheral spatial distribution of the chemical field around the droplet (0 to 2$\pi$ radians) is evaluated at a distance of 10 $\mu$m away from the droplet's periphery, for which the grayscale intensity of the fluorescence micrographs is used to generate the kymograph plots (see figure \ref{fig4}(B-D)). For droplets swimming in just the aqueous TTAB solution, kymograph plot shown in figure \ref{fig4}(B) indicates that while the droplet is in contact with the wall ($t$ $\sim$ 0), the filled micelles are distributed throughout the anterior ($\theta \sim 0$) to the posterior ($\theta \sim \pi$) region around the droplet. Such a distribution of filled micelles is anticipated to not only decelerate the propulsion speed but also impede the detachment of the droplet from the wall, compelling it to glide along the capillary's wall. Nevertheless, over time, droplet detachment was observed, and we will delve into the reasons in subsequent sections. Once the droplet moves away from the wall and swims along the central region of the capillary, filled micelles are primarily distributed in a narrow region on its posterior side. Corresponding grayscale fluorescence micrographs for droplet swimming in these two scenarios are displayed in figures \ref{fig4}(E) and \ref{fig4}(F), respectively.

In presence of PEO, a non-zero, albeit lower compared to the posterior, intensity was measured at the equatorial region (see figure \ref{fig4}(C)), suggesting a more enveloping chemical field around the droplet in this case. The presence of filled micelles at the lateral positions (i.e. $\theta$ = $\pi$/2 and 3$\pi$/2) provides a cushioning effect to the droplet, preventing it from coming into direct contact with the wall due to the negative chemotaxis effect \cite{jin2017chemotaxis}. The corresponding grayscale optical micrograph is shown in figure \ref{fig4}(G). The steady temporal evolution of the kymograph plot confirms the consistent droplet motion along the center line while maintaining a constant separation from the wall. In presence of glycerol, as seen from the optical micrograph (figure \ref{fig4}(H)), the chemical field around the droplet appears distributed non-uniformly from the lateral to the posterior regions. The kymograph plot (figure \ref{fig4}(D)) indicates temporal fluctuations in this chemical field. This behavior is consistent with the previously reported jittery motion due to higher ($\sim$ 3 times) solubilization rate of droplets in presence of 80 wt.$\%$ glycerol \cite{dwivedi2021solute}. The scattered plume of the chemical field, covering lateral angular positions, again acts as a cushion preventing the droplet from approaching the wall. However, due to the temporal fluctuations of this lateral cushioning effect, discernible undulations appear in the trajectory. It was also observed that the lateral plume of chemical field spills to the front of the droplet, see fluorescence micrograph shown in supporting figure S3,  causing a reduction in the droplet's propulsion speed. Nonetheless, the motion remains largely restricted along the capillary axis (+$x$ direction). These characteristics of the chemical field's spatial distribution are quantitatively compared in Figure \ref{fig4}(I), where the angular distribution of I($\theta$) during the axial motion of the swimming droplets in the three cases is illustrated. As expected, for droplets swimming in the aqueous TTAB solution alone, I($\theta$) demonstrates a unimodal normal distribution with a peak located at $\theta \sim \pi$, and tails at 0 and $2\pi$. In the presence of PEO, while maintaining a similar normal distribution, the width increases compared to the former. Upon the addition of glycerol, the distribution shows noticeable angular fluctuations.

\begin{figure*}[ht!]
  {\includegraphics[scale=0.6]{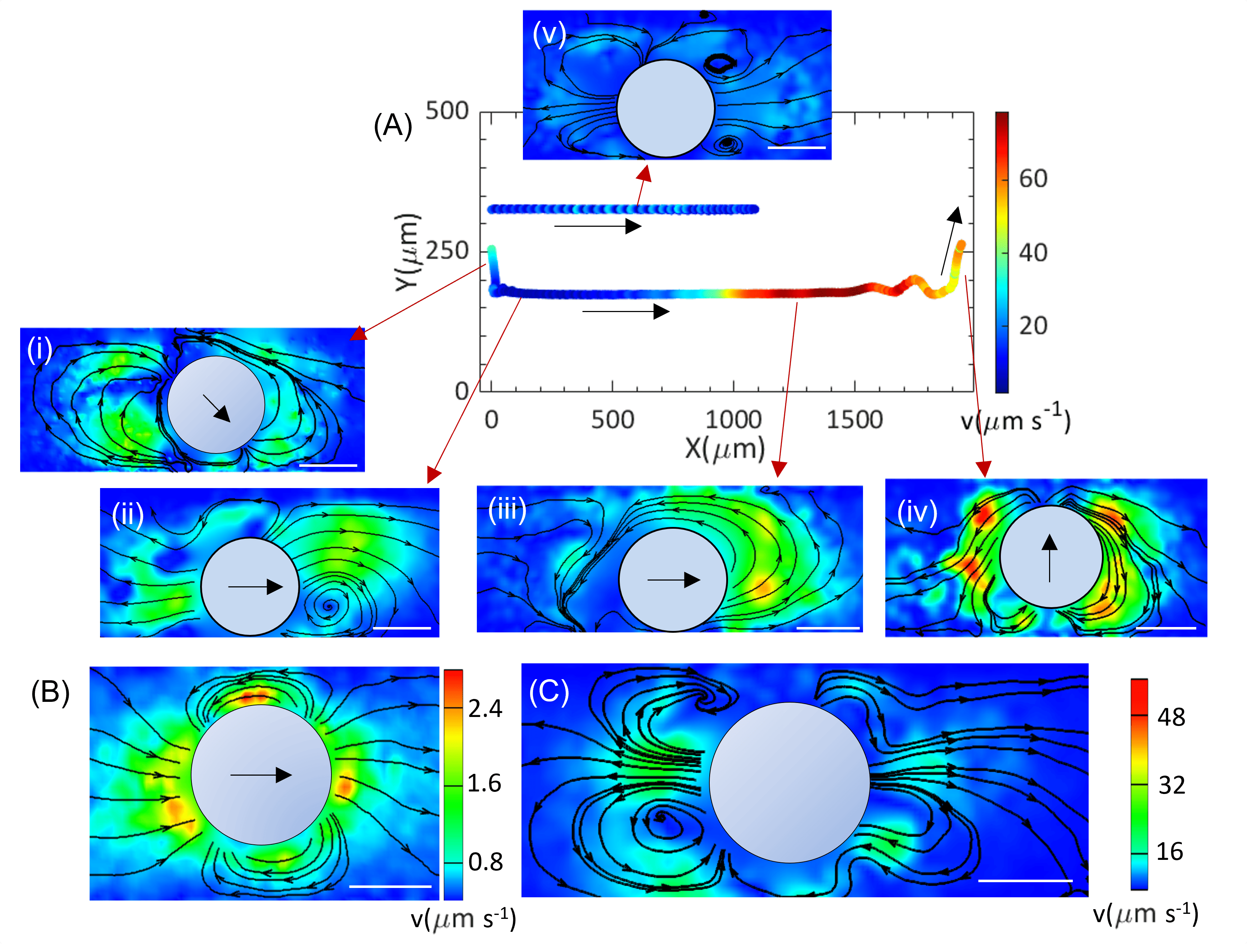}}
\linespread{1.0}
\caption{\small (Color online). (A) Representative trajectory of a 5CB droplet moving along the wall in TTAB solution and (i-v) show the PIV micrographs of fluid flow field around the droplet at different occasions of droplet (i) approaching towards the wall, (ii) motion along the wall, (iii) before detachment with wall, (iv) lateral movement after detachment and (v) rest state. (B, C) PIV micrographs of fluid flow field around the droplet for PEO and glycerol, respectively. Scale bar = 200 $\mu$m. }
\label{fig5}
\end{figure*}

After understanding the spatial distribution of the chemical field around the droplet within the capillary, we proceeded to characterize the surrounding flow field in the droplet's vicinity through PIV experiments. Figure \ref{fig5}(A) illustrates a representative trajectory of an active 5CB droplet swimming in just aqueous TTAB solution, depicting its approach to one of the side walls at $x \sim 10 \mu$m, subsequent movement along the wall, and its eventual detachment at $x \sim 1850 \mu$m. In our experiments, PIV micrographs shown in figure \ref{fig5} demonstrate the local fluid flow field captured in different cases. The asymmetry in flow field surrounding the droplet, figure \ref{fig5}(A(i)), brings it closer to the wall. There, the flow field undergoes further modification, wherein, the streamlines are advected away from the wall (figure \ref{fig5}(A(ii)), which ensures that the droplet glides along the wall. With time once the droplet gains speed, more fluid is brought in vertically towards the wall (figure \ref{fig5}(A(iii)) from the posterior region. Eventually, as the streamlines encourage vertical motion, as seen in figure \ref{fig5}(A(iv)), the droplet detaches. Here, we believe that approach of the droplet towards the wall and subsequent departure is stochastic in nature and originates due to local concentration fluctuations of the available free micelles. In a 2014 study, Li and Ardekani conducted numerical investigations into the hydrodynamic response affecting the swimming behavior of a squirmer near a wall \cite{li2014hydrodynamic}. The study predicts that depending on the prescribed $\beta$ of the squirmer, three distinct types of swimming behaviors can be observed a. swimming away from the wall ($\beta \leq 1$), b. damped oscillations near the wall ($ 2 \leq \beta \leq 5$) and c. bouncing on the wall in cyclic motion ($\beta \geq 7$). In contrast to the fixed $\beta$ in the mentioned numerical study, the observed variation in streamlines around the droplets in our experiments suggests a potential change in $\beta$ as the droplets approach and detach from the wall. While this observed variation in $\beta$ may align with the numerical prediction, it's important to note that verifying this agreement is challenging due to the difficulties in experimentally measuring $\beta$ in close proximity to the wall.

Figure \ref{fig5}(B) depicts the PIV micrograph for a swimming droplet in the presence of PEO in 1D confinement. In contrast to the behavior observed in a 2D Hele-Shaw cell, where the droplet adopts a weak pusher swimming mode, the lateral confinement here results in neutral swimmer-like characteristics. This is corroborated by the measured $\beta$ values, which are close to 0 (refer to supporting figure S2). The reduction in $\beta$ aligns with a more persistent motion within the capillary. Furthermore, during the halt state for droplets swimming in the presence of glycerol, a distorted but somewhat quadrupole-like mode is observed (see figure \ref{fig5}(C)).
\begin{figure*}[ht!]
  {\includegraphics[scale=0.62]{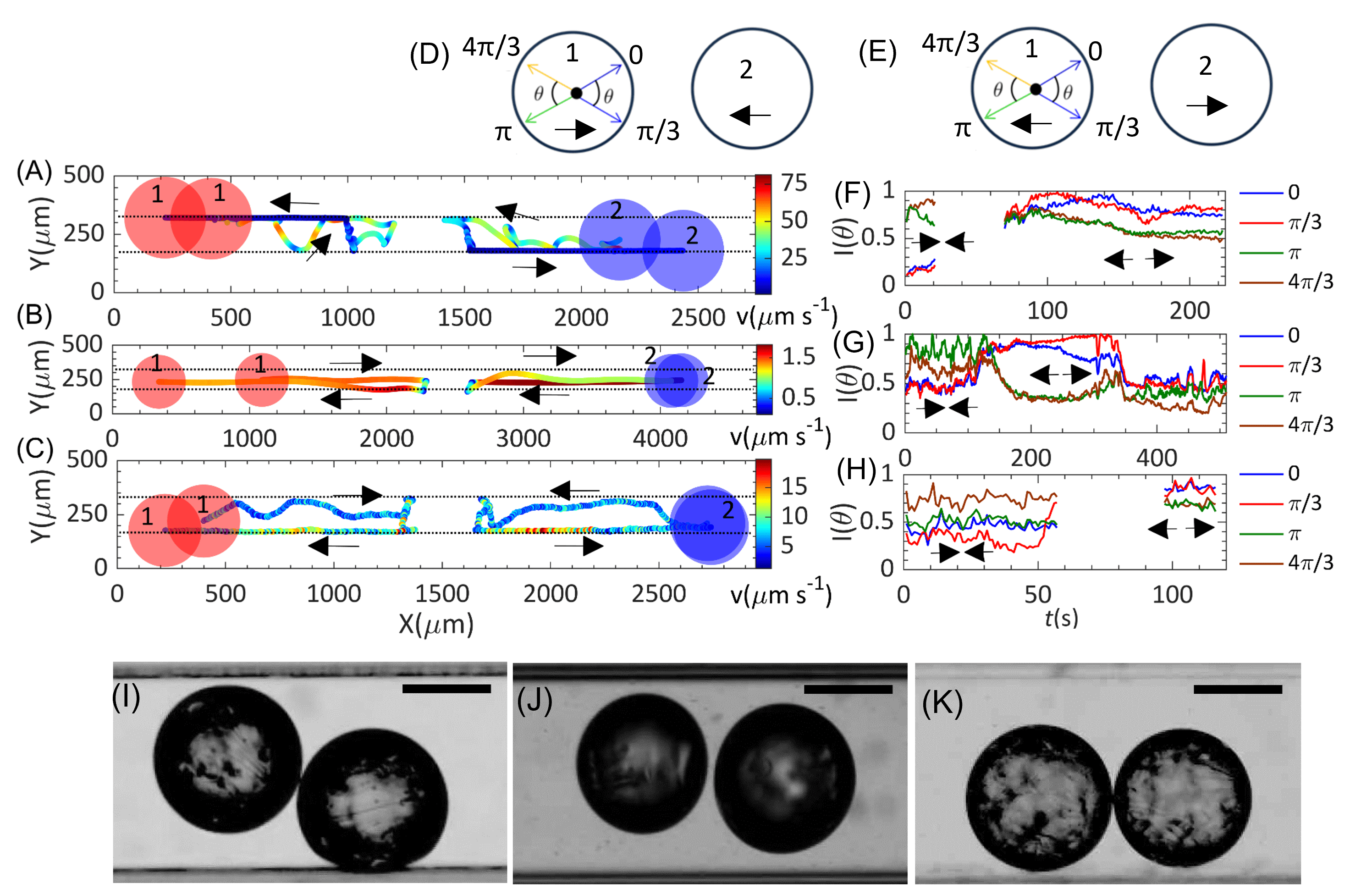}}
\linespread{1.0}
\caption{\small Trajectories (color coded by instantaneous speed) of swimming 5CB droplets after being injected from the opposite ends of the capillary in (A) aqueous TTAB solution with no additive (B) upon adding PEO and (C) upon adding glycerol. Schematic to determine the fluorescence intensity around droplet-1 at (0 and $\pi$/3) and ($\pi$ and 4$\pi$/3) angular positions when droplets (D) approach and (E) scatter. Temporal evolution of fluorescence intensities different angular positions for droplet-1 swimming in (F) aqueous TTAB solution with no additive (G) upon adding PEO and (H) upon adding glycerol. Arrows indicate the droplet's direction of motion. Grayscale micrographs of head-on interactions of droplets at their minimal distance in aqueous TTAB solution with (I) no additive, (J) upon adding PEO, and (K) upon adding glycerol. Scale bar = 200 $\mu$m.}
\label{fig6}
\end{figure*}

\subsection{Pair interactions}

Having gained an understanding of the behavior of individual droplets in quasi 1D confinement, next we investigate the dynamics and the underlying interactions between two 5CB droplets swimming within the quasi-1D confinement of glass capillary. Initially, the droplets were introduced into the capillary from opposite ends (refer to supporting movies S4-S6) resulting in droplet motion towards each other, see figure \ref{fig6}(A-C). In all cases, after initial unrestricted swimming, the droplets decelerate as they approach each other head-on, eventually coming to a stop at their closest point. Subsequently, the droplets reverse their direction of motion and begin moving apart. In order to understand this phenomenon, we carried out fluorescence experiments aimed at examining the distribution of filled micelles. This was achieved by gauging the intensity of the chemical field surrounding droplet-1. Figure \ref{fig6}(F-H) illustrates the fluorescence intensity of the chemical field, represented as $I(\theta)$, recorded at angular positions 0 and $\pi$/3 at the front, and $\pi$ and 4$\pi$/3 at the back, as illustrated in the schematic \ref{fig6}(D, E), while droplet-1 was in motion. In all cases, as anticipated, initially when the droplets were distant from each other, consistently a lower intensity of the chemical field at the front, compared to the back, was measured. Nevertheless, with time as the droplets came closer to each other, filled micelles accumulated at the front of the droplet. As a consequence, the interfacial polarity/asymmetry of the droplets is expected to reduce bringing them to a halt. Over time, the concentration of the interstitial chemical field exceeds the intensity of the chemical field at the back, leading to repulsive interactions that push the droplets apart. Despite inherent variations in the corresponding Péclet number ($Pe$) among the three cases, the observed similarity in pair-wise interactions and subsequent diverging dynamics is intriguing. However, a notable distinction arises in the minimum surface-to-surface distance $d_{min}$ during their closest approach. In the cases of pure water and in presence of glycerol, the droplets seem to touch each other before drifting apart ($d_{min}$ $=$ 0). In contrast, in presence of PEO, the droplets rebound with a minimum distance of approximately 15 µm before separating (see figure \ref{fig6}(I-K)). In 2019, Liperra \textit{et al.} conducted a numerical investigation into the axi-symmetric head-on collision of identical chemically active droplets within a bi-spherical geometry \cite{lippera2020collisions}. The study projected that irrespective of the underlying Péclet number ($Pe$), the approaching droplets would consistently undergo a chemical repulsion due to the accumulation of the chemical field in the intermediate region. This would lead to a gradual deceleration, followed by a change in direction, and eventual re-acceleration as the droplets drifted apart. The research also emphasized that at low $Pe$, the chemical field dominated the hydrodynamic signature of the droplets, causing them to rebound at larger inter-droplet distances. This is in contrast to droplets at higher $Pe$, where stronger hydrodynamic fields enable the droplets to approach more closely before rebounding. Our experimental observations of pair-wise interactions in systems with varying $Pe$ align excellently with this numerical prediction.

\begin{figure*}[ht!]
  {\includegraphics[scale=0.6]{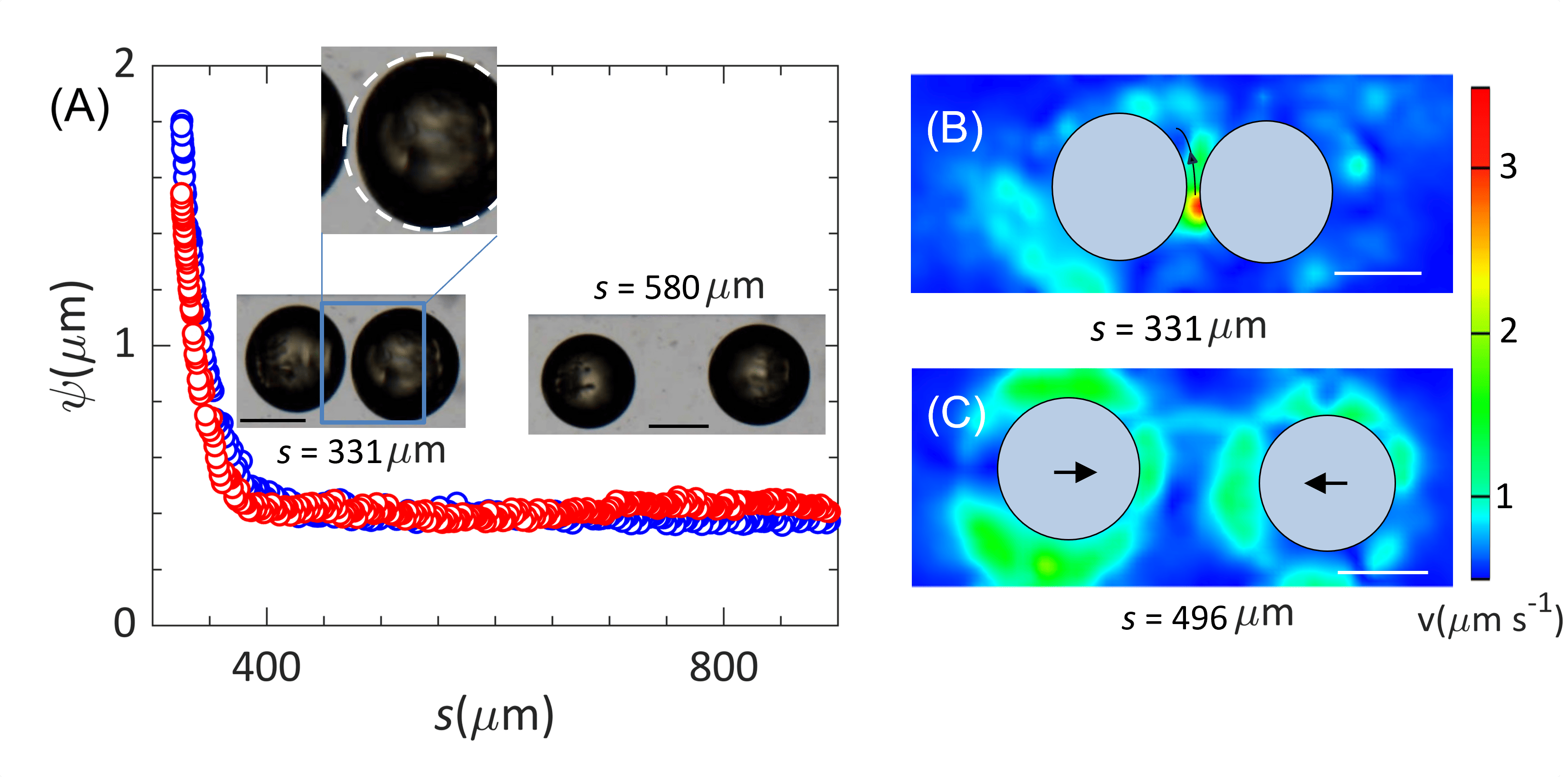}}
\linespread{1.0}
\caption{\small Variation in deformation index ($\psi$) with inter-droplet distance ($s$) for 5CB swimming droplets (A) in aqueous TTAB solution containing 1.125 wt.$\%$ PEO. Insets illustrate the optical micrographs of the droplets both in deformed and undeformed states. (B, C) PIV micrographs of fluid flow field around the droplet for 1.125 wt.$\%$ PEO.  Scale bar: 200 $\mu$m.}
\label{fig7}
\end{figure*}

In case of PEO solution, another intriguing observation is witnessed. During their closest approach, the droplets appear compressed from their facing sides. A decrease in the curvature on anterior regions of both droplets confirmed their shape deformation from their usual spherical shape, see optical micro-graph shown in the inset to figure \ref{fig7}(A)). This deformation was measured in terms of deformation factor $\psi$ as the standard deviation in droplet radius calculated by determining the positions of points on the droplet's periphery with respect to the centroid. Figure \ref{fig7}(A) demonstrates the variation in $\psi$ with inter-droplet distance $s$. For PEO solution, with decreasing $s$, $\psi$ stays nearly zero up to a certain value $\sim$ 400 $\mu$m, and on further decrease in $s$, a sharp rise in $\psi$ is observed. In our recent work, we demonstrated similar shape deformation in an active 5CB droplet swimming in a 2D optical cell filled with high molecular weight polymer solution \cite{dwivedi2023deforming}. Through careful investigation, we argued that a swimming droplet is capable of generating strong fluid flow (convection) around it capable of stretching/compressing the surrounding polymer chains. If the polymer chains are advected faster compared to their equilibrium relaxation time, they develop a residual stress. This competition of time scales is quantified using Deborah number $De$ = $\frac{v\tau}{R}$, where $v$ represents fluid convection, $\tau$ is the polymer relaxation time, and $R$ is the droplet size. At higher $De$, the droplet generates high strain rates ($\frac{v}{R}$) and as a consequence, the unrelaxed polymer chains generate extra normal stresses on the droplet interface resulting in droplet deformation. During the free swimming of droplets in the polymer solution, no droplet deformation is observed, and it is only when two droplets approach each other closely from opposite directions that any deformation is evident. Contrary to our earlier findings, at this instant of maximum deformation, the droplets were the slowest, suggesting associated $De$ be the lowest as well. It is important to highlight that Dwivedi \textit{et al.} used the droplet speed as a measure of fluid-flow to estimate the strain rate experienced by the polymer chains, in the context of an isolated freely swimming droplet. However, when two droplets approach from opposite ends and interact closely, their overlapping plumes of chemical field in the intermediate region generate strong fluid flow which decelerates them significantly. Since, at this instant, droplets are nearly at halt, this strong fluid advection in the intermediate zone is not accurately represented by droplets' individual speeds, instead should be measured using PIV experiments. Similarly, the droplet size ($R$) is not an accurate estimate of length scale over which the fluid convection decays, rather it must be half of the intermediate gap thickness between the droplets' surfaces at their closest approach $L_{min.}$ where their speeds are the lowest. PIV measurements (see figure \ref{fig7}(B,C)) reveal strong fluid flow ($v$ $\sim$ 3.5 $\mu$m s$^{-1}$) in the intermediate region. For $\tau$ $\sim$ 2 s (see Ref.\cite{dwivedi2023deforming}) and an intermediate gap of $L_{min.}$ $\sim$ 9 $\mu$m, it results in an associated $De$ = $\frac{2 v\tau}{L_{min.}}$ $\sim$ 1.5, verifying droplets deformation due to extra normal stresses imparted by the ambient stretched polymer chains. 

\begin{figure*}[ht!]
  {\includegraphics[scale=0.5]{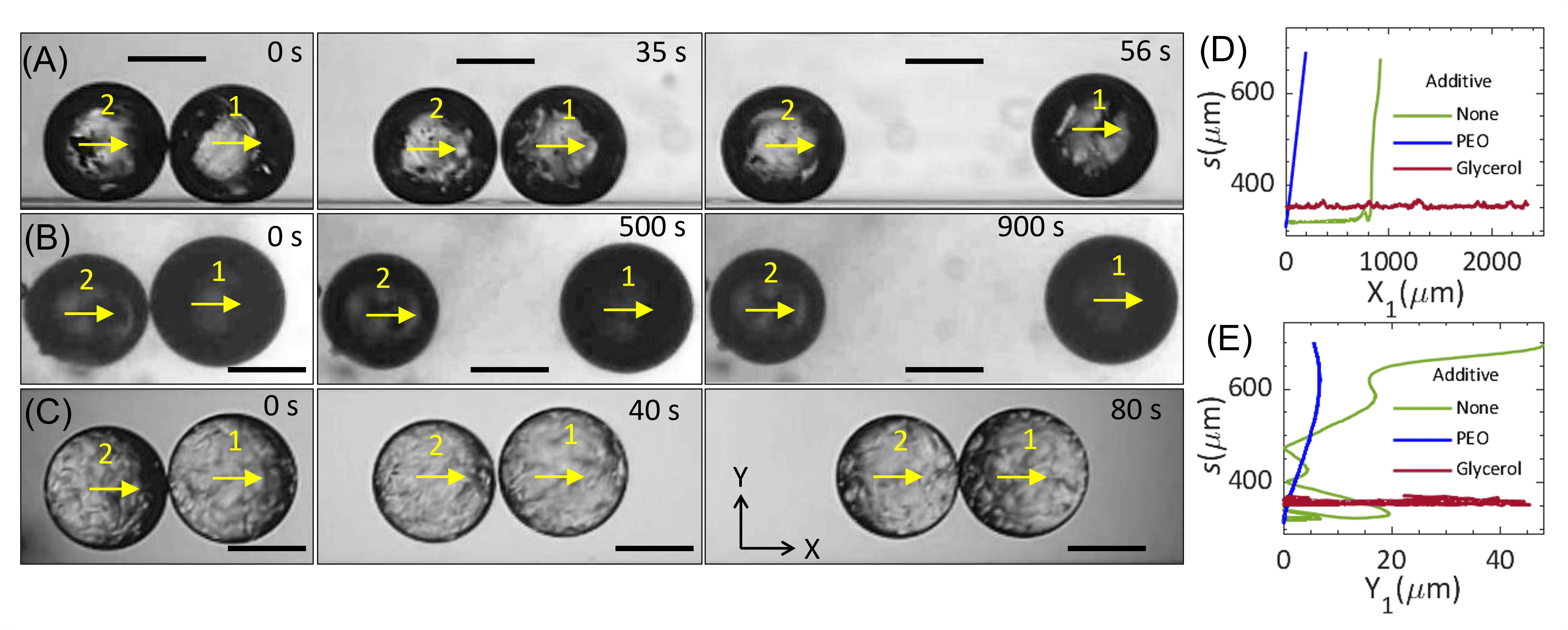}}
\linespread{1.0}
\caption{\small Grayscale optical micrographs of swimming 5CB droplets after being injected from the same end of the capillary in aqueous TTAB solution (A) with no additive, (B) upon adding PEO and (C) upon adding glycerol. Scale bar = 200 $\mu m$. (D) and (E) display the corresponding inter-droplet separation from their centroids with the distance traveled by the leading droplet-1 from its initial position in the $X$ and $Y$ directions, respectively. }
\label{fig8}
\end{figure*}
Subsequently, the droplets were introduced in rapid succession from the same end of the capillary. In Figure \ref{fig8}(A), optical microscopic images taken at different time instances show a pair of droplets moving within a TTAB aqueous solution. Here, droplet-1 takes the lead and droplet-2 appears to follow it, as they move along the wall. With time, when the leading droplet's speed increases, it eventually detaches from the wall, resulting in the discontinuation of the trailing behavior of droplet-2. This dynamic response is captured by the change in the inter-droplet distance ($s$) with the $x$-distance (X$_{1}$) traversed by the leading droplet, measured from its initial position, as illustrated in figure \ref{fig8}(D). Consistent with the random nature of droplet's detachment from the wall, this pair dynamics was observed to vary over time and across different regions within the capillary. In case of PEO solution, no such coupled motion was observed. In fact, a monotonic increase in $s$ with X$_{1}$ (see figure \ref{fig8}(D)) indicates an inherent repulsive interaction between the droplets. On the other hand, in presence of glycerol, as shown in the optical micrographs captured at 0 s, 40 s, and 80 s in figure \ref{fig8}(C), the droplets mostly moved together. The brown line in figure \ref{fig8}(D) depicts negligible variation in $s$ with X$_1$. In fact, it was observed that the droplet-2 chases the droplet-1 throughout the length of the capillary (see supporting movie S7). Further, we plotted the inter-droplet distance ($s$) with the $y$-distance (Y$_{1}$) traversed by the leading droplet, measured from its initial position, as illustrated in figure \ref{fig8}(E). In just water, significant variations in $s$ are observed with fluctuations in Y$_{1}$, whereas, in presence of glycerol $s$ remains unchanged. This difference suggests that the coupled motion of droplets observed in case of just water is not actually a chasing behavior, instead, their forced movement along the wall results in an impression of droplet-2 chasing the droplet-1. 

\begin{figure}[ht!]
  {\includegraphics[width=8.4cm,height=9.4cm]{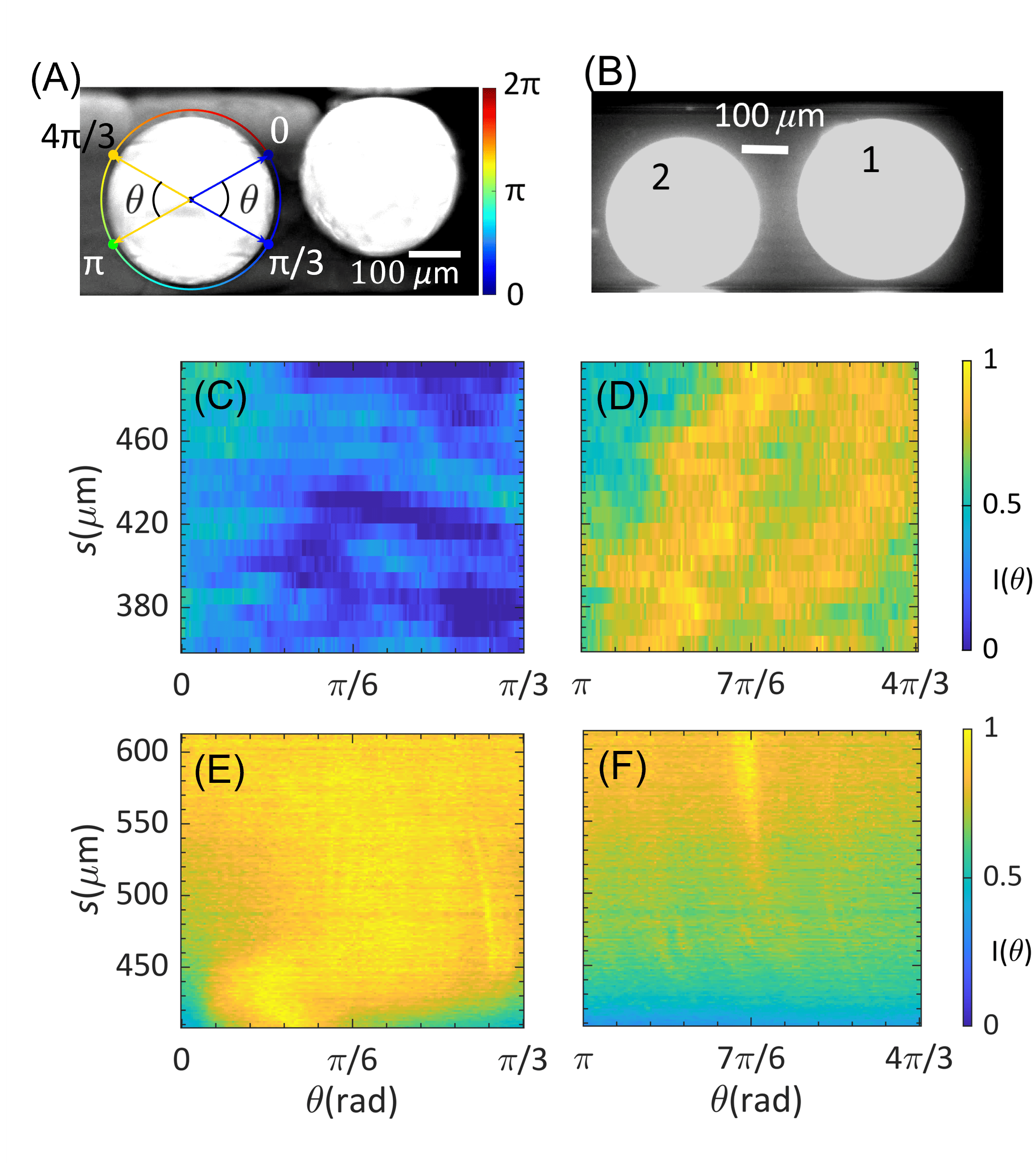}}
\linespread{1.0}
\caption{\small Grayscale fluorescence micrograph of a pair of 5CB droplets highlighting the intermediate plume of chemical field while swimming (left to right) in aqueous TTAB solution in presence of (A) glycerol and (B) PEO. Kymograph plots of the chasing droplet around the (C) anterior and (D) posterior regimes. (E) and (F) represent the kymographs for the chemical field around the rear droplet's anterior and posterior regimes, respectively, for a pair of 5CB swimming droplets in aqueous TTAB solution containing PEO.}
\label{fig9}
\end{figure}

To comprehend the chasing behavior observed in glycerol solution, we conducted fluorescence experiments to assess the chemical field's intensity at the anterior (0 to $\pi$/3) and the posterior ($\pi$ to 4$\pi$/3) regions of the trailing droplet-2, with varying the inter-droplet distance $s$ (see figure \ref{fig9}(A)). As depicted in figure \ref{fig9}(C, D) the posterior region exhibits a notably higher presence of the chemical field compared to the anterior region. We suspect that during the approach of the trailing droplet, due to higher Péclet ($Pe$), the combined advection effects in the intermediate region, sweep the filled micelles from that region towards the rear end of the trailing droplet. This revised asymmetry in the distribution of the chemical field results in a more pronounced flow field in the negative $x$-direction supporting the forward motion (+$x$-direction) of the trailing droplet toward the leading droplet (see PIV micrograph shown in supporting figure S4). 

In case of PEO solution, as demonstrated in figure \ref{fig9}(E, F), a contrasting scenario unfolds. Here, at the anterior region of the rear droplet, a higher concentration of the chemical field accumulates between the two droplets (figure \ref{fig9}(E, F)), generating a chemotactic repulsion between the droplets. As a consequence, the separation between the droplets increases with time. The underlying reason for this disparity in chemical field distribution is due to the difference in the associated Péclet ($Pe$) numbers. In pure water and glycerol solution, higher $Pe$ values promote the distribution of the chemical field away from the droplet through stronger advection. However, this is not the case in presence of PEO, where owing to smaller $Pe$ values, the distribution of micelles primarily occurs through diffusion, which results in a localized accumulation in the intermediate region between the droplets.

The similarity of the repulsive interactions between the pair of 5CB droplets in presence of PEO during their head-on and head-tail interactions emphasizes the primary influence of chemical field-mediated interactions due to the low $Pe$. This aligns with the previously documented more concentrated distribution of the chemical field around the droplets in solutions involving macromolecules (as noted in \cite{dwivedi2023mode}). However, the varying behavior of pair-interactions in higher $Pe$ systems (in presence of glycerol, and in just water) indicates differences in their hydrodynamic interactions during their head-on and head-tail approaches, resulting in distinct distributions of the chemical fields. In the head-tail approach, it seems that the leading droplet's ability to move rightward to gather free surfactants creates more robust advection currents around it, preventing the buildup of filled micelles in the space between. However, during the head-on approach, as neither droplet has the opportunity to gather free surfactants from either side, the currents weaken, leading to the accumulation of filled micelles in the interstitial region. The PIV micrograph shown in supporting figure S4 offers evidence of higher convection in the head-tail approach as compared to the head-on approach. Nevertheless, understanding these orientation-specific interactions requires further exploration through a dedicated 2D investigation, which we intend to undertake in near future. In 2020, Liperra \textit{et al.} reported on the asymmetric head-on collisions between different sized droplets with large $Pe$ difference and predicted a tailing behavior wherein the droplet with larger $Pe$ chases the droplet with lower $Pe$ \cite{lippera2020bouncing}. This study further highlights the complexity of the pair-wise interactions of droplets which underscores the need for more elaborate experimental investigations.

%%%%%%%%%%%%%%%%%

%The spherical symmetry of a droplet also breaks in the viscoelastic medium which is imposed by the elastic stress.\cite{dwivedi2023deforming,poryles2018encapsulation} 

\section{Conclusions}
The current study presents a comprehensive investigation into the micellar solubilization-driven self-propelled motion of 5CB droplets within an aqueous surfactant (TTAB) solution in a square capillary. Our exploration aims to understand how the physical one-dimensional confinement within the capillary affects the swimming behavior of isolated 5CB droplets and their pair interactions at varying Péclet numbers ($Pe$). We introduce macromolecular (high molecular weight PEO) and molecular (glycerol) solutes to adjust the underlying $Pe$. Employing detailed characterization techniques for both hydrodynamic field (PIV) and chemical field (fluorescence) surrounding the droplets, we illustrate how the one-dimensional confinement provided by the capillary walls alters the distribution of the chemical field around the droplet, resulting in a notable difference in droplet swimming compared to a 2D geometry. This effect is distinctly observed in the three cases where the strength of micelle advection to diffusion, represented by the Péclet number ($Pe$), differs. In PEO and glycerol solutions, an accumulation of the chemical field between the droplet and the capillary wall leads to droplet motion away from the wall. Conversely, in the absence of any additive to the surfactant solution, irregular motion with occasional gliding along the wall and subsequent detachment is observed.

Regarding pair-wise interactions, head-on approaching droplets scatter in all scenarios. In the case of PEO, at the closest point of approach, droplets deform from their typical spherical shape due to the increase in the local $De$ associated with intermediate fluid-flow. In glycerol, when moving in the same direction, droplets exhibit attraction resulting in chasing behavior. However, when approaching from opposite ends, they scatter after reaching the closest point. Our observation of scattering in all scenarios is explained based on the distribution of the chemical field during their interaction. However, the chasing behavior in glycerol is expected due to the stronger advection that influences the distribution of the chemical field.

\section{acknowledgement}
We acknowledge the funding received by Department of Science and Technology (SR/FST/ETII055/2013), Science and Engineering Research Board (Grant numbers SB/S2/RJN105/2017 and CRG/2022/003763), Department of Science and Technology, India.

%\end{acknowledgement}

%\begin{suppinfo}

\section{Supplementary movies}
S1: Complex behavior of isolated 5CB droplet in a square capillary in 6 wt$\%$ TTAB aqueous solution, shown at 10 X the actual speed.

S2: 5CB droplet in a square capillary self-propel in a straight line by maintaining the separation with boundaries in 1.125 wt.$\%$ PEO aqueous TTAB solution, shown at 150 X the actual speed.

S3: Jittery behavior of 5CB droplet avoiding contact with capillary boundaries in 80 wt$\%$ glycerol aqueous TTAB solution, shown at 15 X the actual speed.

S4: Repulsion of 5CB droplets upon injection of opposite ends of capillary in TTAB aqueous solution. Movie is shown at 10 X the actual speed.

S5: Repulsion of 5CB droplets upon injection of opposite ends of capillary in 1.125 wt.$\%$ PEO aqueous TTAB solution. Movie is shown at 150 X the actual speed.

S6: Repulsion of 5CB droplets upon injection of opposite ends of capillary in 80 wt.$\%$ glycerol aqueous TTAB solution. Movie is shown at 30 X the actual speed.

S7: Chasing behavior of 5CB droplets upon injection of one side of the capillary containing 80 wt.$\%$ glycerol aqueous TTAB solution, shown at 10 X the actual speed.
%Mean square displacement and normalized velocity autocorrelation function curves of isolated active JPs, trajectories of active JP pairs partaking in different collision scenarios.

%\end{suppinfo}

\bibliography{arxiv}

\begin{thebibliography}{48}%
\makeatletter
\providecommand \@ifxundefined [1]{%
 \@ifx{#1\undefined}
}%
\providecommand \@ifnum [1]{%
 \ifnum #1\expandafter \@firstoftwo
 \else \expandafter \@secondoftwo
 \fi
}%
\providecommand \@ifx [1]{%
 \ifx #1\expandafter \@firstoftwo
 \else \expandafter \@secondoftwo
 \fi
}%
\providecommand \natexlab [1]{#1}%
\providecommand \enquote  [1]{``#1''}%
\providecommand \bibnamefont  [1]{#1}%
\providecommand \bibfnamefont [1]{#1}%
\providecommand \citenamefont [1]{#1}%
\providecommand \href@noop [0]{\@secondoftwo}%
\providecommand \href [0]{\begingroup \@sanitize@url \@href}%
\providecommand \@href[1]{\@@startlink{#1}\@@href}%
\providecommand \@@href[1]{\endgroup#1\@@endlink}%
\providecommand \@sanitize@url [0]{\catcode `\\12\catcode `\$12\catcode
  `\&12\catcode `\#12\catcode `\^12\catcode `\_12\catcode `\%12\relax}%
\providecommand \@@startlink[1]{}%
\providecommand \@@endlink[0]{}%
\providecommand \url  [0]{\begingroup\@sanitize@url \@url }%
\providecommand \@url [1]{\endgroup\@href {#1}{\urlprefix }}%
\providecommand \urlprefix  [0]{URL }%
\providecommand \Eprint [0]{\href }%
\providecommand \doibase [0]{https://doi.org/}%
\providecommand \selectlanguage [0]{\@gobble}%
\providecommand \bibinfo  [0]{\@secondoftwo}%
\providecommand \bibfield  [0]{\@secondoftwo}%
\providecommand \translation [1]{[#1]}%
\providecommand \BibitemOpen [0]{}%
\providecommand \bibitemStop [0]{}%
\providecommand \bibitemNoStop [0]{.\EOS\space}%
\providecommand \EOS [0]{\spacefactor3000\relax}%
\providecommand \BibitemShut  [1]{\csname bibitem#1\endcsname}%
\let\auto@bib@innerbib\@empty
%</preamble>
\bibitem [{\citenamefont {Purcell}(1977)}]{purcell1977life}%
  \BibitemOpen
  \bibfield  {author} {\bibinfo {author} {\bibfnamefont {E.~M.}\ \bibnamefont
  {Purcell}},\ }\href@noop {} {\bibfield  {journal} {\bibinfo  {journal}
  {American journal of physics}\ }\textbf {\bibinfo {volume} {45}},\ \bibinfo
  {pages} {3} (\bibinfo {year} {1977})}\BibitemShut {NoStop}%
\bibitem [{\citenamefont {Z{\"o}ttl}\ and\ \citenamefont
  {Yeomans}(2019)}]{zottl2019enhanced}%
  \BibitemOpen
  \bibfield  {author} {\bibinfo {author} {\bibfnamefont {A.}~\bibnamefont
  {Z{\"o}ttl}}\ and\ \bibinfo {author} {\bibfnamefont {J.~M.}\ \bibnamefont
  {Yeomans}},\ }\href@noop {} {\bibfield  {journal} {\bibinfo  {journal}
  {Nature Physics}\ }\textbf {\bibinfo {volume} {15}},\ \bibinfo {pages} {554}
  (\bibinfo {year} {2019})}\BibitemShut {NoStop}%
\bibitem [{\citenamefont {Noselli}\ \emph {et~al.}(2019)\citenamefont
  {Noselli}, \citenamefont {Beran}, \citenamefont {Arroyo},\ and\ \citenamefont
  {DeSimone}}]{noselli2019swimming}%
  \BibitemOpen
  \bibfield  {author} {\bibinfo {author} {\bibfnamefont {G.}~\bibnamefont
  {Noselli}}, \bibinfo {author} {\bibfnamefont {A.}~\bibnamefont {Beran}},
  \bibinfo {author} {\bibfnamefont {M.}~\bibnamefont {Arroyo}},\ and\ \bibinfo
  {author} {\bibfnamefont {A.}~\bibnamefont {DeSimone}},\ }\href@noop {}
  {\bibfield  {journal} {\bibinfo  {journal} {Nature physics}\ }\textbf
  {\bibinfo {volume} {15}},\ \bibinfo {pages} {496} (\bibinfo {year}
  {2019})}\BibitemShut {NoStop}%
\bibitem [{\citenamefont {Xu}\ \emph {et~al.}(2018)\citenamefont {Xu},
  \citenamefont {Medina-S{\'a}nchez}, \citenamefont {Magdanz}, \citenamefont
  {Schwarz}, \citenamefont {Hebenstreit},\ and\ \citenamefont
  {Schmidt}}]{xu2018sperm}%
  \BibitemOpen
  \bibfield  {author} {\bibinfo {author} {\bibfnamefont {H.}~\bibnamefont
  {Xu}}, \bibinfo {author} {\bibfnamefont {M.}~\bibnamefont
  {Medina-S{\'a}nchez}}, \bibinfo {author} {\bibfnamefont {V.}~\bibnamefont
  {Magdanz}}, \bibinfo {author} {\bibfnamefont {L.}~\bibnamefont {Schwarz}},
  \bibinfo {author} {\bibfnamefont {F.}~\bibnamefont {Hebenstreit}},\ and\
  \bibinfo {author} {\bibfnamefont {O.~G.}\ \bibnamefont {Schmidt}},\
  }\href@noop {} {\bibfield  {journal} {\bibinfo  {journal} {ACS nano}\
  }\textbf {\bibinfo {volume} {12}},\ \bibinfo {pages} {327} (\bibinfo {year}
  {2018})}\BibitemShut {NoStop}%
\bibitem [{\citenamefont {Gao}\ and\ \citenamefont
  {Wang}(2014)}]{gao2014environmental}%
  \BibitemOpen
  \bibfield  {author} {\bibinfo {author} {\bibfnamefont {W.}~\bibnamefont
  {Gao}}\ and\ \bibinfo {author} {\bibfnamefont {J.}~\bibnamefont {Wang}},\
  }\href@noop {} {\bibfield  {journal} {\bibinfo  {journal} {Acs Nano}\
  }\textbf {\bibinfo {volume} {8}},\ \bibinfo {pages} {3170} (\bibinfo {year}
  {2014})}\BibitemShut {NoStop}%
\bibitem [{\citenamefont {Jurado-S{\'a}nchez}\ and\ \citenamefont
  {Escarpa}(2017)}]{jurado2017janus}%
  \BibitemOpen
  \bibfield  {author} {\bibinfo {author} {\bibfnamefont {B.}~\bibnamefont
  {Jurado-S{\'a}nchez}}\ and\ \bibinfo {author} {\bibfnamefont
  {A.}~\bibnamefont {Escarpa}},\ }\href@noop {} {\bibfield  {journal} {\bibinfo
   {journal} {Electroanalysis}\ }\textbf {\bibinfo {volume} {29}},\ \bibinfo
  {pages} {14} (\bibinfo {year} {2017})}\BibitemShut {NoStop}%
\bibitem [{\citenamefont {W{\"u}rger}(2015)}]{wurger2015self}%
  \BibitemOpen
  \bibfield  {author} {\bibinfo {author} {\bibfnamefont {A.}~\bibnamefont
  {W{\"u}rger}},\ }\href@noop {} {\bibfield  {journal} {\bibinfo  {journal}
  {Physical review letters}\ }\textbf {\bibinfo {volume} {115}},\ \bibinfo
  {pages} {188304} (\bibinfo {year} {2015})}\BibitemShut {NoStop}%
\bibitem [{\citenamefont {Zhou}\ \emph {et~al.}(2018)\citenamefont {Zhou},
  \citenamefont {Zhang}, \citenamefont {Tang},\ and\ \citenamefont
  {Wang}}]{zhou2018photochemically}%
  \BibitemOpen
  \bibfield  {author} {\bibinfo {author} {\bibfnamefont {C.}~\bibnamefont
  {Zhou}}, \bibinfo {author} {\bibfnamefont {H.}~\bibnamefont {Zhang}},
  \bibinfo {author} {\bibfnamefont {J.}~\bibnamefont {Tang}},\ and\ \bibinfo
  {author} {\bibfnamefont {W.}~\bibnamefont {Wang}},\ }\href@noop {} {\bibfield
   {journal} {\bibinfo  {journal} {Langmuir}\ }\textbf {\bibinfo {volume}
  {34}},\ \bibinfo {pages} {3289} (\bibinfo {year} {2018})}\BibitemShut
  {NoStop}%
\bibitem [{\citenamefont {Jiang}\ \emph {et~al.}(2010)\citenamefont {Jiang},
  \citenamefont {Yoshinaga},\ and\ \citenamefont {Sano}}]{jiang2010active}%
  \BibitemOpen
  \bibfield  {author} {\bibinfo {author} {\bibfnamefont {H.-R.}\ \bibnamefont
  {Jiang}}, \bibinfo {author} {\bibfnamefont {N.}~\bibnamefont {Yoshinaga}},\
  and\ \bibinfo {author} {\bibfnamefont {M.}~\bibnamefont {Sano}},\ }\href@noop
  {} {\bibfield  {journal} {\bibinfo  {journal} {Physical review letters}\
  }\textbf {\bibinfo {volume} {105}},\ \bibinfo {pages} {268302} (\bibinfo
  {year} {2010})}\BibitemShut {NoStop}%
\bibitem [{\citenamefont {Cao}\ \emph {et~al.}(2021)\citenamefont {Cao},
  \citenamefont {Shao}, \citenamefont {Wu}, \citenamefont {Song}, \citenamefont
  {De~Martino}, \citenamefont {Pijpers}, \citenamefont {Friedrich},
  \citenamefont {Abdelmohsen}, \citenamefont {Williams},\ and\ \citenamefont
  {van Hest}}]{cao2021photoactivated}%
  \BibitemOpen
  \bibfield  {author} {\bibinfo {author} {\bibfnamefont {S.}~\bibnamefont
  {Cao}}, \bibinfo {author} {\bibfnamefont {J.}~\bibnamefont {Shao}}, \bibinfo
  {author} {\bibfnamefont {H.}~\bibnamefont {Wu}}, \bibinfo {author}
  {\bibfnamefont {S.}~\bibnamefont {Song}}, \bibinfo {author} {\bibfnamefont
  {M.~T.}\ \bibnamefont {De~Martino}}, \bibinfo {author} {\bibfnamefont
  {I.~A.}\ \bibnamefont {Pijpers}}, \bibinfo {author} {\bibfnamefont
  {H.}~\bibnamefont {Friedrich}}, \bibinfo {author} {\bibfnamefont {L.~K.}\
  \bibnamefont {Abdelmohsen}}, \bibinfo {author} {\bibfnamefont {D.~S.}\
  \bibnamefont {Williams}},\ and\ \bibinfo {author} {\bibfnamefont {J.~C.}\
  \bibnamefont {van Hest}},\ }\href@noop {} {\bibfield  {journal} {\bibinfo
  {journal} {Nature communications}\ }\textbf {\bibinfo {volume} {12}},\
  \bibinfo {pages} {2077} (\bibinfo {year} {2021})}\BibitemShut {NoStop}%
\bibitem [{\citenamefont {Dou}\ \emph {et~al.}(2016)\citenamefont {Dou},
  \citenamefont {Cartier}, \citenamefont {Fei}, \citenamefont {Pandey},
  \citenamefont {Razavi}, \citenamefont {Kretzschmar},\ and\ \citenamefont
  {Bishop}}]{dou2016directed}%
  \BibitemOpen
  \bibfield  {author} {\bibinfo {author} {\bibfnamefont {Y.}~\bibnamefont
  {Dou}}, \bibinfo {author} {\bibfnamefont {C.~A.}\ \bibnamefont {Cartier}},
  \bibinfo {author} {\bibfnamefont {W.}~\bibnamefont {Fei}}, \bibinfo {author}
  {\bibfnamefont {S.}~\bibnamefont {Pandey}}, \bibinfo {author} {\bibfnamefont
  {S.}~\bibnamefont {Razavi}}, \bibinfo {author} {\bibfnamefont
  {I.}~\bibnamefont {Kretzschmar}},\ and\ \bibinfo {author} {\bibfnamefont
  {K.~J.}\ \bibnamefont {Bishop}},\ }\href@noop {} {\bibfield  {journal}
  {\bibinfo  {journal} {Langmuir}\ }\textbf {\bibinfo {volume} {32}},\ \bibinfo
  {pages} {13167} (\bibinfo {year} {2016})}\BibitemShut {NoStop}%
\bibitem [{\citenamefont {Gangwal}\ \emph {et~al.}(2008)\citenamefont
  {Gangwal}, \citenamefont {Cayre}, \citenamefont {Bazant},\ and\ \citenamefont
  {Velev}}]{gangwal2008induced}%
  \BibitemOpen
  \bibfield  {author} {\bibinfo {author} {\bibfnamefont {S.}~\bibnamefont
  {Gangwal}}, \bibinfo {author} {\bibfnamefont {O.~J.}\ \bibnamefont {Cayre}},
  \bibinfo {author} {\bibfnamefont {M.~Z.}\ \bibnamefont {Bazant}},\ and\
  \bibinfo {author} {\bibfnamefont {O.~D.}\ \bibnamefont {Velev}},\ }\href@noop
  {} {\bibfield  {journal} {\bibinfo  {journal} {Physical review letters}\
  }\textbf {\bibinfo {volume} {100}},\ \bibinfo {pages} {058302} (\bibinfo
  {year} {2008})}\BibitemShut {NoStop}%
\bibitem [{\citenamefont {Dwivedi}\ \emph
  {et~al.}(2021{\natexlab{a}})\citenamefont {Dwivedi}, \citenamefont {Si},
  \citenamefont {Pillai},\ and\ \citenamefont {Mangal}}]{dwivedi2021solute}%
  \BibitemOpen
  \bibfield  {author} {\bibinfo {author} {\bibfnamefont {P.}~\bibnamefont
  {Dwivedi}}, \bibinfo {author} {\bibfnamefont {B.~R.}\ \bibnamefont {Si}},
  \bibinfo {author} {\bibfnamefont {D.}~\bibnamefont {Pillai}},\ and\ \bibinfo
  {author} {\bibfnamefont {R.}~\bibnamefont {Mangal}},\ }\href@noop {}
  {\bibfield  {journal} {\bibinfo  {journal} {Physics of Fluids}\ }\textbf
  {\bibinfo {volume} {33}} (\bibinfo {year} {2021}{\natexlab{a}})}\BibitemShut
  {NoStop}%
\bibitem [{\citenamefont {Izzet}\ \emph {et~al.}(2020)\citenamefont {Izzet},
  \citenamefont {Moerman}, \citenamefont {Gross}, \citenamefont {Groenewold},
  \citenamefont {Hollingsworth}, \citenamefont {Bibette},\ and\ \citenamefont
  {Brujic}}]{izzet2020tunable}%
  \BibitemOpen
  \bibfield  {author} {\bibinfo {author} {\bibfnamefont {A.}~\bibnamefont
  {Izzet}}, \bibinfo {author} {\bibfnamefont {P.~G.}\ \bibnamefont {Moerman}},
  \bibinfo {author} {\bibfnamefont {P.}~\bibnamefont {Gross}}, \bibinfo
  {author} {\bibfnamefont {J.}~\bibnamefont {Groenewold}}, \bibinfo {author}
  {\bibfnamefont {A.~D.}\ \bibnamefont {Hollingsworth}}, \bibinfo {author}
  {\bibfnamefont {J.}~\bibnamefont {Bibette}},\ and\ \bibinfo {author}
  {\bibfnamefont {J.}~\bibnamefont {Brujic}},\ }\href@noop {} {\bibfield
  {journal} {\bibinfo  {journal} {Physical Review X}\ }\textbf {\bibinfo
  {volume} {10}},\ \bibinfo {pages} {021035} (\bibinfo {year}
  {2020})}\BibitemShut {NoStop}%
\bibitem [{\citenamefont {Izri}\ \emph {et~al.}(2014)\citenamefont {Izri},
  \citenamefont {Van Der~Linden}, \citenamefont {Michelin},\ and\ \citenamefont
  {Dauchot}}]{izri2014self}%
  \BibitemOpen
  \bibfield  {author} {\bibinfo {author} {\bibfnamefont {Z.}~\bibnamefont
  {Izri}}, \bibinfo {author} {\bibfnamefont {M.~N.}\ \bibnamefont {Van
  Der~Linden}}, \bibinfo {author} {\bibfnamefont {S.}~\bibnamefont
  {Michelin}},\ and\ \bibinfo {author} {\bibfnamefont {O.}~\bibnamefont
  {Dauchot}},\ }\href@noop {} {\bibfield  {journal} {\bibinfo  {journal}
  {Physical review letters}\ }\textbf {\bibinfo {volume} {113}},\ \bibinfo
  {pages} {248302} (\bibinfo {year} {2014})}\BibitemShut {NoStop}%
\bibitem [{\citenamefont {Meredith}\ \emph {et~al.}(2022)\citenamefont
  {Meredith}, \citenamefont {Castonguay}, \citenamefont {Chiu}, \citenamefont
  {Brooks}, \citenamefont {Moerman}, \citenamefont {Torab}, \citenamefont
  {Wong}, \citenamefont {Sen}, \citenamefont {Velegol},\ and\ \citenamefont
  {Zarzar}}]{meredith2022chemical}%
  \BibitemOpen
  \bibfield  {author} {\bibinfo {author} {\bibfnamefont {C.~H.}\ \bibnamefont
  {Meredith}}, \bibinfo {author} {\bibfnamefont {A.~C.}\ \bibnamefont
  {Castonguay}}, \bibinfo {author} {\bibfnamefont {Y.-J.}\ \bibnamefont
  {Chiu}}, \bibinfo {author} {\bibfnamefont {A.~M.}\ \bibnamefont {Brooks}},
  \bibinfo {author} {\bibfnamefont {P.~G.}\ \bibnamefont {Moerman}}, \bibinfo
  {author} {\bibfnamefont {P.}~\bibnamefont {Torab}}, \bibinfo {author}
  {\bibfnamefont {P.~K.}\ \bibnamefont {Wong}}, \bibinfo {author}
  {\bibfnamefont {A.}~\bibnamefont {Sen}}, \bibinfo {author} {\bibfnamefont
  {D.}~\bibnamefont {Velegol}},\ and\ \bibinfo {author} {\bibfnamefont {L.~D.}\
  \bibnamefont {Zarzar}},\ }\href@noop {} {\bibfield  {journal} {\bibinfo
  {journal} {Matter}\ }\textbf {\bibinfo {volume} {5}},\ \bibinfo {pages} {616}
  (\bibinfo {year} {2022})}\BibitemShut {NoStop}%
\bibitem [{\citenamefont {Peddireddy}\ \emph {et~al.}(2012)\citenamefont
  {Peddireddy}, \citenamefont {Kumar}, \citenamefont {Thutupalli},
  \citenamefont {Herminghaus},\ and\ \citenamefont
  {Bahr}}]{peddireddy2012solubilization}%
  \BibitemOpen
  \bibfield  {author} {\bibinfo {author} {\bibfnamefont {K.}~\bibnamefont
  {Peddireddy}}, \bibinfo {author} {\bibfnamefont {P.}~\bibnamefont {Kumar}},
  \bibinfo {author} {\bibfnamefont {S.}~\bibnamefont {Thutupalli}}, \bibinfo
  {author} {\bibfnamefont {S.}~\bibnamefont {Herminghaus}},\ and\ \bibinfo
  {author} {\bibfnamefont {C.}~\bibnamefont {Bahr}},\ }\href@noop {} {\bibfield
   {journal} {\bibinfo  {journal} {Langmuir}\ }\textbf {\bibinfo {volume}
  {28}},\ \bibinfo {pages} {12426} (\bibinfo {year} {2012})}\BibitemShut
  {NoStop}%
\bibitem [{\citenamefont {Thakur}\ \emph {et~al.}(2006)\citenamefont {Thakur},
  \citenamefont {Kumar}, \citenamefont {Madhusudana},\ and\ \citenamefont
  {Pullarkat}}]{thakur2006self}%
  \BibitemOpen
  \bibfield  {author} {\bibinfo {author} {\bibfnamefont {S.}~\bibnamefont
  {Thakur}}, \bibinfo {author} {\bibfnamefont {P.~S.}\ \bibnamefont {Kumar}},
  \bibinfo {author} {\bibfnamefont {N.}~\bibnamefont {Madhusudana}},\ and\
  \bibinfo {author} {\bibfnamefont {P.~A.}\ \bibnamefont {Pullarkat}},\
  }\href@noop {} {\bibfield  {journal} {\bibinfo  {journal} {Physical review
  letters}\ }\textbf {\bibinfo {volume} {97}},\ \bibinfo {pages} {115701}
  (\bibinfo {year} {2006})}\BibitemShut {NoStop}%
\bibitem [{\citenamefont {Kumar}\ \emph {et~al.}(2021)\citenamefont {Kumar},
  \citenamefont {Borkar},\ and\ \citenamefont {Dayal}}]{kumar2021fast}%
  \BibitemOpen
  \bibfield  {author} {\bibinfo {author} {\bibfnamefont {D.~J.~P.}\
  \bibnamefont {Kumar}}, \bibinfo {author} {\bibfnamefont {C.}~\bibnamefont
  {Borkar}},\ and\ \bibinfo {author} {\bibfnamefont {P.}~\bibnamefont
  {Dayal}},\ }\href@noop {} {\bibfield  {journal} {\bibinfo  {journal}
  {Langmuir}\ }\textbf {\bibinfo {volume} {37}},\ \bibinfo {pages} {12586}
  (\bibinfo {year} {2021})}\BibitemShut {NoStop}%
\bibitem [{\citenamefont {Morozov}\ and\ \citenamefont
  {Michelin}(2019)}]{morozov2019nonlinear}%
  \BibitemOpen
  \bibfield  {author} {\bibinfo {author} {\bibfnamefont {M.}~\bibnamefont
  {Morozov}}\ and\ \bibinfo {author} {\bibfnamefont {S.}~\bibnamefont
  {Michelin}},\ }\href@noop {} {\bibfield  {journal} {\bibinfo  {journal} {The
  Journal of chemical physics}\ }\textbf {\bibinfo {volume} {150}} (\bibinfo
  {year} {2019})}\BibitemShut {NoStop}%
\bibitem [{\citenamefont {Dwivedi}\ \emph
  {et~al.}(2023{\natexlab{a}})\citenamefont {Dwivedi}, \citenamefont
  {Shrivastava}, \citenamefont {Pillai},\ and\ \citenamefont
  {Mangal}}]{dwivedi2023mode}%
  \BibitemOpen
  \bibfield  {author} {\bibinfo {author} {\bibfnamefont {P.}~\bibnamefont
  {Dwivedi}}, \bibinfo {author} {\bibfnamefont {A.}~\bibnamefont
  {Shrivastava}}, \bibinfo {author} {\bibfnamefont {D.}~\bibnamefont
  {Pillai}},\ and\ \bibinfo {author} {\bibfnamefont {R.}~\bibnamefont
  {Mangal}},\ }\href@noop {} {\bibfield  {journal} {\bibinfo  {journal} {Soft
  Matter}\ }\textbf {\bibinfo {volume} {19}},\ \bibinfo {pages} {4099}
  (\bibinfo {year} {2023}{\natexlab{a}})}\BibitemShut {NoStop}%
\bibitem [{\citenamefont {Suda}\ \emph {et~al.}(2021)\citenamefont {Suda},
  \citenamefont {Suda}, \citenamefont {Ohmura},\ and\ \citenamefont
  {Ichikawa}}]{suda2021straight}%
  \BibitemOpen
  \bibfield  {author} {\bibinfo {author} {\bibfnamefont {S.}~\bibnamefont
  {Suda}}, \bibinfo {author} {\bibfnamefont {T.}~\bibnamefont {Suda}}, \bibinfo
  {author} {\bibfnamefont {T.}~\bibnamefont {Ohmura}},\ and\ \bibinfo {author}
  {\bibfnamefont {M.}~\bibnamefont {Ichikawa}},\ }\href@noop {} {\bibfield
  {journal} {\bibinfo  {journal} {Physical Review Letters}\ }\textbf {\bibinfo
  {volume} {127}},\ \bibinfo {pages} {088005} (\bibinfo {year}
  {2021})}\BibitemShut {NoStop}%
\bibitem [{\citenamefont {Kr{\"u}ger}\ \emph {et~al.}(2016)\citenamefont
  {Kr{\"u}ger}, \citenamefont {Kl{\"o}s}, \citenamefont {Bahr},\ and\
  \citenamefont {Maass}}]{kruger2016curling}%
  \BibitemOpen
  \bibfield  {author} {\bibinfo {author} {\bibfnamefont {C.}~\bibnamefont
  {Kr{\"u}ger}}, \bibinfo {author} {\bibfnamefont {G.}~\bibnamefont
  {Kl{\"o}s}}, \bibinfo {author} {\bibfnamefont {C.}~\bibnamefont {Bahr}},\
  and\ \bibinfo {author} {\bibfnamefont {C.~C.}\ \bibnamefont {Maass}},\
  }\href@noop {} {\bibfield  {journal} {\bibinfo  {journal} {Physical review
  letters}\ }\textbf {\bibinfo {volume} {117}},\ \bibinfo {pages} {048003}
  (\bibinfo {year} {2016})}\BibitemShut {NoStop}%
\bibitem [{\citenamefont {Suga}\ \emph {et~al.}(2018)\citenamefont {Suga},
  \citenamefont {Suda}, \citenamefont {Ichikawa},\ and\ \citenamefont
  {Kimura}}]{suga2018self}%
  \BibitemOpen
  \bibfield  {author} {\bibinfo {author} {\bibfnamefont {M.}~\bibnamefont
  {Suga}}, \bibinfo {author} {\bibfnamefont {S.}~\bibnamefont {Suda}}, \bibinfo
  {author} {\bibfnamefont {M.}~\bibnamefont {Ichikawa}},\ and\ \bibinfo
  {author} {\bibfnamefont {Y.}~\bibnamefont {Kimura}},\ }\href@noop {}
  {\bibfield  {journal} {\bibinfo  {journal} {Physical Review E}\ }\textbf
  {\bibinfo {volume} {97}},\ \bibinfo {pages} {062703} (\bibinfo {year}
  {2018})}\BibitemShut {NoStop}%
\bibitem [{\citenamefont {Hokmabad}\ \emph {et~al.}(2021)\citenamefont
  {Hokmabad}, \citenamefont {Dey}, \citenamefont {Jalaal}, \citenamefont
  {Mohanty}, \citenamefont {Almukambetova}, \citenamefont {Baldwin},
  \citenamefont {Lohse},\ and\ \citenamefont {Maass}}]{hokmabad2021emergence}%
  \BibitemOpen
  \bibfield  {author} {\bibinfo {author} {\bibfnamefont {B.~V.}\ \bibnamefont
  {Hokmabad}}, \bibinfo {author} {\bibfnamefont {R.}~\bibnamefont {Dey}},
  \bibinfo {author} {\bibfnamefont {M.}~\bibnamefont {Jalaal}}, \bibinfo
  {author} {\bibfnamefont {D.}~\bibnamefont {Mohanty}}, \bibinfo {author}
  {\bibfnamefont {M.}~\bibnamefont {Almukambetova}}, \bibinfo {author}
  {\bibfnamefont {K.~A.}\ \bibnamefont {Baldwin}}, \bibinfo {author}
  {\bibfnamefont {D.}~\bibnamefont {Lohse}},\ and\ \bibinfo {author}
  {\bibfnamefont {C.~C.}\ \bibnamefont {Maass}},\ }\href@noop {} {\bibfield
  {journal} {\bibinfo  {journal} {Physical review X}\ }\textbf {\bibinfo
  {volume} {11}},\ \bibinfo {pages} {011043} (\bibinfo {year}
  {2021})}\BibitemShut {NoStop}%
\bibitem [{\citenamefont {Jin}\ \emph {et~al.}(2017)\citenamefont {Jin},
  \citenamefont {Kr{\"u}ger},\ and\ \citenamefont {Maass}}]{jin2017chemotaxis}%
  \BibitemOpen
  \bibfield  {author} {\bibinfo {author} {\bibfnamefont {C.}~\bibnamefont
  {Jin}}, \bibinfo {author} {\bibfnamefont {C.}~\bibnamefont {Kr{\"u}ger}},\
  and\ \bibinfo {author} {\bibfnamefont {C.~C.}\ \bibnamefont {Maass}},\
  }\href@noop {} {\bibfield  {journal} {\bibinfo  {journal} {Proceedings of the
  National Academy of Sciences}\ }\textbf {\bibinfo {volume} {114}},\ \bibinfo
  {pages} {5089} (\bibinfo {year} {2017})}\BibitemShut {NoStop}%
\bibitem [{\citenamefont {Dwivedi}\ \emph
  {et~al.}(2021{\natexlab{b}})\citenamefont {Dwivedi}, \citenamefont
  {Shrivastava}, \citenamefont {Pillai},\ and\ \citenamefont
  {Mangal}}]{dwivedi2021rheotaxis}%
  \BibitemOpen
  \bibfield  {author} {\bibinfo {author} {\bibfnamefont {P.}~\bibnamefont
  {Dwivedi}}, \bibinfo {author} {\bibfnamefont {A.}~\bibnamefont
  {Shrivastava}}, \bibinfo {author} {\bibfnamefont {D.}~\bibnamefont
  {Pillai}},\ and\ \bibinfo {author} {\bibfnamefont {R.}~\bibnamefont
  {Mangal}},\ }\href@noop {} {\bibfield  {journal} {\bibinfo  {journal}
  {Physics of Fluids}\ }\textbf {\bibinfo {volume} {33}} (\bibinfo {year}
  {2021}{\natexlab{b}})}\BibitemShut {NoStop}%
\bibitem [{\citenamefont {Dey}\ \emph {et~al.}(2022)\citenamefont {Dey},
  \citenamefont {Buness}, \citenamefont {Hokmabad}, \citenamefont {Jin},\ and\
  \citenamefont {Maass}}]{dey2022oscillatory}%
  \BibitemOpen
  \bibfield  {author} {\bibinfo {author} {\bibfnamefont {R.}~\bibnamefont
  {Dey}}, \bibinfo {author} {\bibfnamefont {C.~M.}\ \bibnamefont {Buness}},
  \bibinfo {author} {\bibfnamefont {B.~V.}\ \bibnamefont {Hokmabad}}, \bibinfo
  {author} {\bibfnamefont {C.}~\bibnamefont {Jin}},\ and\ \bibinfo {author}
  {\bibfnamefont {C.~C.}\ \bibnamefont {Maass}},\ }\href@noop {} {\bibfield
  {journal} {\bibinfo  {journal} {Nature communications}\ }\textbf {\bibinfo
  {volume} {13}},\ \bibinfo {pages} {2952} (\bibinfo {year}
  {2022})}\BibitemShut {NoStop}%
\bibitem [{\citenamefont {Dwivedi}\ \emph
  {et~al.}(2023{\natexlab{b}})\citenamefont {Dwivedi}, \citenamefont
  {Shrivastava}, \citenamefont {Pillai}, \citenamefont {Tiwari},\ and\
  \citenamefont {Mangal}}]{dwivedi2023deforming}%
  \BibitemOpen
  \bibfield  {author} {\bibinfo {author} {\bibfnamefont {P.}~\bibnamefont
  {Dwivedi}}, \bibinfo {author} {\bibfnamefont {A.}~\bibnamefont
  {Shrivastava}}, \bibinfo {author} {\bibfnamefont {D.}~\bibnamefont {Pillai}},
  \bibinfo {author} {\bibfnamefont {N.}~\bibnamefont {Tiwari}},\ and\ \bibinfo
  {author} {\bibfnamefont {R.}~\bibnamefont {Mangal}},\ }\href@noop {}
  {\bibfield  {journal} {\bibinfo  {journal} {Soft Matter}\ }\textbf {\bibinfo
  {volume} {19}},\ \bibinfo {pages} {3783} (\bibinfo {year}
  {2023}{\natexlab{b}})}\BibitemShut {NoStop}%
\bibitem [{\citenamefont {Meredith}\ \emph {et~al.}(2020)\citenamefont
  {Meredith}, \citenamefont {Moerman}, \citenamefont {Groenewold},
  \citenamefont {Chiu}, \citenamefont {Kegel}, \citenamefont {van Blaaderen},\
  and\ \citenamefont {Zarzar}}]{meredith2020predator}%
  \BibitemOpen
  \bibfield  {author} {\bibinfo {author} {\bibfnamefont {C.~H.}\ \bibnamefont
  {Meredith}}, \bibinfo {author} {\bibfnamefont {P.~G.}\ \bibnamefont
  {Moerman}}, \bibinfo {author} {\bibfnamefont {J.}~\bibnamefont {Groenewold}},
  \bibinfo {author} {\bibfnamefont {Y.-J.}\ \bibnamefont {Chiu}}, \bibinfo
  {author} {\bibfnamefont {W.~K.}\ \bibnamefont {Kegel}}, \bibinfo {author}
  {\bibfnamefont {A.}~\bibnamefont {van Blaaderen}},\ and\ \bibinfo {author}
  {\bibfnamefont {L.~D.}\ \bibnamefont {Zarzar}},\ }\href@noop {} {\bibfield
  {journal} {\bibinfo  {journal} {Nature Chemistry}\ }\textbf {\bibinfo
  {volume} {12}},\ \bibinfo {pages} {1136} (\bibinfo {year}
  {2020})}\BibitemShut {NoStop}%
\bibitem [{\citenamefont {Hokmabad}\ \emph
  {et~al.}(2022{\natexlab{a}})\citenamefont {Hokmabad}, \citenamefont
  {Nishide}, \citenamefont {Ramesh}, \citenamefont {Kr{\"u}ger},\ and\
  \citenamefont {Maass}}]{hokmabad2022spontaneously}%
  \BibitemOpen
  \bibfield  {author} {\bibinfo {author} {\bibfnamefont {B.~V.}\ \bibnamefont
  {Hokmabad}}, \bibinfo {author} {\bibfnamefont {A.}~\bibnamefont {Nishide}},
  \bibinfo {author} {\bibfnamefont {P.}~\bibnamefont {Ramesh}}, \bibinfo
  {author} {\bibfnamefont {C.}~\bibnamefont {Kr{\"u}ger}},\ and\ \bibinfo
  {author} {\bibfnamefont {C.~C.}\ \bibnamefont {Maass}},\ }\href@noop {}
  {\bibfield  {journal} {\bibinfo  {journal} {Soft matter}\ }\textbf {\bibinfo
  {volume} {18}},\ \bibinfo {pages} {2731} (\bibinfo {year}
  {2022}{\natexlab{a}})}\BibitemShut {NoStop}%
\bibitem [{\citenamefont {Thutupalli}\ \emph {et~al.}(2018)\citenamefont
  {Thutupalli}, \citenamefont {Geyer}, \citenamefont {Singh}, \citenamefont
  {Adhikari},\ and\ \citenamefont {Stone}}]{thutupalli2018flow}%
  \BibitemOpen
  \bibfield  {author} {\bibinfo {author} {\bibfnamefont {S.}~\bibnamefont
  {Thutupalli}}, \bibinfo {author} {\bibfnamefont {D.}~\bibnamefont {Geyer}},
  \bibinfo {author} {\bibfnamefont {R.}~\bibnamefont {Singh}}, \bibinfo
  {author} {\bibfnamefont {R.}~\bibnamefont {Adhikari}},\ and\ \bibinfo
  {author} {\bibfnamefont {H.~A.}\ \bibnamefont {Stone}},\ }\href@noop {}
  {\bibfield  {journal} {\bibinfo  {journal} {Proceedings of the National
  Academy of Sciences}\ }\textbf {\bibinfo {volume} {115}},\ \bibinfo {pages}
  {5403} (\bibinfo {year} {2018})}\BibitemShut {NoStop}%
\bibitem [{\citenamefont {Hokmabad}\ \emph
  {et~al.}(2022{\natexlab{b}})\citenamefont {Hokmabad}, \citenamefont
  {Agudo-Canalejo}, \citenamefont {Saha}, \citenamefont {Golestanian},\ and\
  \citenamefont {Maass}}]{hokmabad2022chemotactic}%
  \BibitemOpen
  \bibfield  {author} {\bibinfo {author} {\bibfnamefont {B.~V.}\ \bibnamefont
  {Hokmabad}}, \bibinfo {author} {\bibfnamefont {J.}~\bibnamefont
  {Agudo-Canalejo}}, \bibinfo {author} {\bibfnamefont {S.}~\bibnamefont
  {Saha}}, \bibinfo {author} {\bibfnamefont {R.}~\bibnamefont {Golestanian}},\
  and\ \bibinfo {author} {\bibfnamefont {C.~C.}\ \bibnamefont {Maass}},\
  }\href@noop {} {\bibfield  {journal} {\bibinfo  {journal} {Proceedings of the
  National Academy of Sciences}\ }\textbf {\bibinfo {volume} {119}},\ \bibinfo
  {pages} {e2122269119} (\bibinfo {year} {2022}{\natexlab{b}})}\BibitemShut
  {NoStop}%
\bibitem [{\citenamefont {Liu}\ \emph {et~al.}(2014)\citenamefont {Liu},
  \citenamefont {Breuer},\ and\ \citenamefont {Powers}}]{liu2014propulsion}%
  \BibitemOpen
  \bibfield  {author} {\bibinfo {author} {\bibfnamefont {B.}~\bibnamefont
  {Liu}}, \bibinfo {author} {\bibfnamefont {K.~S.}\ \bibnamefont {Breuer}},\
  and\ \bibinfo {author} {\bibfnamefont {T.~R.}\ \bibnamefont {Powers}},\
  }\href@noop {} {\bibfield  {journal} {\bibinfo  {journal} {Physics of
  Fluids}\ }\textbf {\bibinfo {volume} {26}} (\bibinfo {year}
  {2014})}\BibitemShut {NoStop}%
\bibitem [{\citenamefont {Acemoglu}\ and\ \citenamefont
  {Yesilyurt}(2014)}]{acemoglu2014effects}%
  \BibitemOpen
  \bibfield  {author} {\bibinfo {author} {\bibfnamefont {A.}~\bibnamefont
  {Acemoglu}}\ and\ \bibinfo {author} {\bibfnamefont {S.}~\bibnamefont
  {Yesilyurt}},\ }\href@noop {} {\bibfield  {journal} {\bibinfo  {journal}
  {Biophysical journal}\ }\textbf {\bibinfo {volume} {106}},\ \bibinfo {pages}
  {1537} (\bibinfo {year} {2014})}\BibitemShut {NoStop}%
\bibitem [{\citenamefont {Wu}\ \emph {et~al.}(2015)\citenamefont {Wu},
  \citenamefont {Thi{\'e}baud}, \citenamefont {Hu}, \citenamefont {Farutin},
  \citenamefont {Rafai}, \citenamefont {Lai}, \citenamefont {Peyla},\ and\
  \citenamefont {Misbah}}]{wu2015amoeboid}%
  \BibitemOpen
  \bibfield  {author} {\bibinfo {author} {\bibfnamefont {H.}~\bibnamefont
  {Wu}}, \bibinfo {author} {\bibfnamefont {M.}~\bibnamefont {Thi{\'e}baud}},
  \bibinfo {author} {\bibfnamefont {W.-F.}\ \bibnamefont {Hu}}, \bibinfo
  {author} {\bibfnamefont {A.}~\bibnamefont {Farutin}}, \bibinfo {author}
  {\bibfnamefont {S.}~\bibnamefont {Rafai}}, \bibinfo {author} {\bibfnamefont
  {M.-C.}\ \bibnamefont {Lai}}, \bibinfo {author} {\bibfnamefont
  {P.}~\bibnamefont {Peyla}},\ and\ \bibinfo {author} {\bibfnamefont
  {C.}~\bibnamefont {Misbah}},\ }\href@noop {} {\bibfield  {journal} {\bibinfo
  {journal} {Physical Review E}\ }\textbf {\bibinfo {volume} {92}},\ \bibinfo
  {pages} {050701} (\bibinfo {year} {2015})}\BibitemShut {NoStop}%
\bibitem [{\citenamefont {Mozaffari}\ \emph {et~al.}(2016)\citenamefont
  {Mozaffari}, \citenamefont {Sharifi-Mood}, \citenamefont {Koplik},\ and\
  \citenamefont {Maldarelli}}]{mozaffari2016self}%
  \BibitemOpen
  \bibfield  {author} {\bibinfo {author} {\bibfnamefont {A.}~\bibnamefont
  {Mozaffari}}, \bibinfo {author} {\bibfnamefont {N.}~\bibnamefont
  {Sharifi-Mood}}, \bibinfo {author} {\bibfnamefont {J.}~\bibnamefont
  {Koplik}},\ and\ \bibinfo {author} {\bibfnamefont {C.}~\bibnamefont
  {Maldarelli}},\ }\href@noop {} {\bibfield  {journal} {\bibinfo  {journal}
  {Physics of Fluids}\ }\textbf {\bibinfo {volume} {28}} (\bibinfo {year}
  {2016})}\BibitemShut {NoStop}%
\bibitem [{\citenamefont {Uspal}\ \emph {et~al.}(2015)\citenamefont {Uspal},
  \citenamefont {Popescu}, \citenamefont {Dietrich},\ and\ \citenamefont
  {Tasinkevych}}]{uspal2015self}%
  \BibitemOpen
  \bibfield  {author} {\bibinfo {author} {\bibfnamefont {W.}~\bibnamefont
  {Uspal}}, \bibinfo {author} {\bibfnamefont {M.~N.}\ \bibnamefont {Popescu}},
  \bibinfo {author} {\bibfnamefont {S.}~\bibnamefont {Dietrich}},\ and\
  \bibinfo {author} {\bibfnamefont {M.}~\bibnamefont {Tasinkevych}},\
  }\href@noop {} {\bibfield  {journal} {\bibinfo  {journal} {Soft matter}\
  }\textbf {\bibinfo {volume} {11}},\ \bibinfo {pages} {434} (\bibinfo {year}
  {2015})}\BibitemShut {NoStop}%
\bibitem [{\citenamefont {Ibrahim}\ and\ \citenamefont
  {Liverpool}(2016)}]{ibrahim2016walls}%
  \BibitemOpen
  \bibfield  {author} {\bibinfo {author} {\bibfnamefont {Y.}~\bibnamefont
  {Ibrahim}}\ and\ \bibinfo {author} {\bibfnamefont {T.~B.}\ \bibnamefont
  {Liverpool}},\ }\href@noop {} {\bibfield  {journal} {\bibinfo  {journal} {The
  European Physical Journal Special Topics}\ }\textbf {\bibinfo {volume}
  {225}},\ \bibinfo {pages} {1843} (\bibinfo {year} {2016})}\BibitemShut
  {NoStop}%
\bibitem [{\citenamefont {Desai}\ and\ \citenamefont
  {Michelin}(2022)}]{desai2022steady}%
  \BibitemOpen
  \bibfield  {author} {\bibinfo {author} {\bibfnamefont {N.}~\bibnamefont
  {Desai}}\ and\ \bibinfo {author} {\bibfnamefont {S.}~\bibnamefont
  {Michelin}},\ }\href@noop {} {\bibfield  {journal} {\bibinfo  {journal}
  {Bulletin of the American Physical Society}\ } (\bibinfo {year}
  {2022})}\BibitemShut {NoStop}%
\bibitem [{\citenamefont {Desai}\ and\ \citenamefont
  {Michelin}(2021)}]{desai2021instability}%
  \BibitemOpen
  \bibfield  {author} {\bibinfo {author} {\bibfnamefont {N.}~\bibnamefont
  {Desai}}\ and\ \bibinfo {author} {\bibfnamefont {S.}~\bibnamefont
  {Michelin}},\ }\href@noop {} {\bibfield  {journal} {\bibinfo  {journal}
  {Physical Review Fluids}\ }\textbf {\bibinfo {volume} {6}},\ \bibinfo {pages}
  {114103} (\bibinfo {year} {2021})}\BibitemShut {NoStop}%
\bibitem [{\citenamefont {Lippera}\ \emph
  {et~al.}(2020{\natexlab{a}})\citenamefont {Lippera}, \citenamefont {Morozov},
  \citenamefont {Benzaquen},\ and\ \citenamefont
  {Michelin}}]{lippera2020collisions}%
  \BibitemOpen
  \bibfield  {author} {\bibinfo {author} {\bibfnamefont {K.}~\bibnamefont
  {Lippera}}, \bibinfo {author} {\bibfnamefont {M.}~\bibnamefont {Morozov}},
  \bibinfo {author} {\bibfnamefont {M.}~\bibnamefont {Benzaquen}},\ and\
  \bibinfo {author} {\bibfnamefont {S.}~\bibnamefont {Michelin}},\ }\href@noop
  {} {\bibfield  {journal} {\bibinfo  {journal} {Journal of Fluid Mechanics}\
  }\textbf {\bibinfo {volume} {886}},\ \bibinfo {pages} {A17} (\bibinfo {year}
  {2020}{\natexlab{a}})}\BibitemShut {NoStop}%
\bibitem [{\citenamefont {de~Blois}\ \emph {et~al.}(2021)\citenamefont
  {de~Blois}, \citenamefont {Bertin}, \citenamefont {Suda}, \citenamefont
  {Ichikawa}, \citenamefont {Reyssat},\ and\ \citenamefont
  {Dauchot}}]{de2021swimming}%
  \BibitemOpen
  \bibfield  {author} {\bibinfo {author} {\bibfnamefont {C.}~\bibnamefont
  {de~Blois}}, \bibinfo {author} {\bibfnamefont {V.}~\bibnamefont {Bertin}},
  \bibinfo {author} {\bibfnamefont {S.}~\bibnamefont {Suda}}, \bibinfo {author}
  {\bibfnamefont {M.}~\bibnamefont {Ichikawa}}, \bibinfo {author}
  {\bibfnamefont {M.}~\bibnamefont {Reyssat}},\ and\ \bibinfo {author}
  {\bibfnamefont {O.}~\bibnamefont {Dauchot}},\ }\href@noop {} {\bibfield
  {journal} {\bibinfo  {journal} {Soft matter}\ }\textbf {\bibinfo {volume}
  {17}},\ \bibinfo {pages} {6646} (\bibinfo {year} {2021})}\BibitemShut
  {NoStop}%
\bibitem [{\citenamefont {Son}\ \emph {et~al.}(2015)\citenamefont {Son},
  \citenamefont {Brumley},\ and\ \citenamefont {Stocker}}]{son2015live}%
  \BibitemOpen
  \bibfield  {author} {\bibinfo {author} {\bibfnamefont {K.}~\bibnamefont
  {Son}}, \bibinfo {author} {\bibfnamefont {D.~R.}\ \bibnamefont {Brumley}},\
  and\ \bibinfo {author} {\bibfnamefont {R.}~\bibnamefont {Stocker}},\
  }\href@noop {} {\bibfield  {journal} {\bibinfo  {journal} {Nature Reviews
  Microbiology}\ }\textbf {\bibinfo {volume} {13}},\ \bibinfo {pages} {761}
  (\bibinfo {year} {2015})}\BibitemShut {NoStop}%
\bibitem [{\citenamefont {Thielicke}\ and\ \citenamefont
  {Sonntag}(2021)}]{thielicke2021particle}%
  \BibitemOpen
  \bibfield  {author} {\bibinfo {author} {\bibfnamefont {W.}~\bibnamefont
  {Thielicke}}\ and\ \bibinfo {author} {\bibfnamefont {R.}~\bibnamefont
  {Sonntag}},\ }\href@noop {} {\bibfield  {journal} {\bibinfo  {journal}
  {Journal of Open Research Software}\ }\textbf {\bibinfo {volume} {9}}
  (\bibinfo {year} {2021})}\BibitemShut {NoStop}%
\bibitem [{\citenamefont {Dwivedi}\ \emph {et~al.}(2022)\citenamefont
  {Dwivedi}, \citenamefont {Pillai},\ and\ \citenamefont
  {Mangal}}]{dwivedi2022self}%
  \BibitemOpen
  \bibfield  {author} {\bibinfo {author} {\bibfnamefont {P.}~\bibnamefont
  {Dwivedi}}, \bibinfo {author} {\bibfnamefont {D.}~\bibnamefont {Pillai}},\
  and\ \bibinfo {author} {\bibfnamefont {R.}~\bibnamefont {Mangal}},\
  }\href@noop {} {\bibfield  {journal} {\bibinfo  {journal} {Current Opinion in
  Colloid \& Interface Science}\ ,\ \bibinfo {pages} {101614}} (\bibinfo {year}
  {2022})}\BibitemShut {NoStop}%
\bibitem [{\citenamefont {Li}\ and\ \citenamefont
  {Ardekani}(2014)}]{li2014hydrodynamic}%
  \BibitemOpen
  \bibfield  {author} {\bibinfo {author} {\bibfnamefont {G.-J.}\ \bibnamefont
  {Li}}\ and\ \bibinfo {author} {\bibfnamefont {A.~M.}\ \bibnamefont
  {Ardekani}},\ }\href@noop {} {\bibfield  {journal} {\bibinfo  {journal}
  {Physical Review E}\ }\textbf {\bibinfo {volume} {90}},\ \bibinfo {pages}
  {013010} (\bibinfo {year} {2014})}\BibitemShut {NoStop}%
\bibitem [{\citenamefont {Lippera}\ \emph
  {et~al.}(2020{\natexlab{b}})\citenamefont {Lippera}, \citenamefont
  {Benzaquen},\ and\ \citenamefont {Michelin}}]{lippera2020bouncing}%
  \BibitemOpen
  \bibfield  {author} {\bibinfo {author} {\bibfnamefont {K.}~\bibnamefont
  {Lippera}}, \bibinfo {author} {\bibfnamefont {M.}~\bibnamefont {Benzaquen}},\
  and\ \bibinfo {author} {\bibfnamefont {S.}~\bibnamefont {Michelin}},\
  }\href@noop {} {\bibfield  {journal} {\bibinfo  {journal} {Physical Review
  Fluids}\ }\textbf {\bibinfo {volume} {5}},\ \bibinfo {pages} {032201}
  (\bibinfo {year} {2020}{\natexlab{b}})}\BibitemShut {NoStop}%
\end{thebibliography}%

%\begin{singlespace}

\end{document}